\documentclass[lettersize,journal]{IEEEtran}
\usepackage{amsmath,amsfonts}
\usepackage{algorithmic}
\usepackage{algorithm}
\usepackage{array}
\usepackage[caption=false,font=normalsize,labelfont=sf,textfont=sf]{subfig}
\usepackage{textcomp}
\usepackage{stfloats}
\usepackage{url}
\usepackage{verbatim}
\usepackage{graphicx}
\usepackage{cite}

\usepackage{bbm}
\usepackage{multirow}

\newcommand{\ba}{\begin{array}}
\newcommand{\ea}{\end{array}}
\newcommand{\be}{\begin{displaymath}}
\newcommand{\ee}{\end{displaymath}}
\newcommand{\ben}{\begin{equation}}
\newcommand{\een}{\end{equation}}
\newcommand{\bena}{\begin{eqnarray}}
\newcommand{\eena}{\end{eqnarray}}
\newcommand{\beqa}{\begin{eqnarray*}}
\newcommand{\enqa}{\end{eqnarray*}}

\newcommand{\bc}{\begin{center}}
\newcommand{\ec}{\end{center}}
\newcommand{\bi}{\begin{itemize}}
\newcommand{\ei}{\end{itemize}}
\newcommand{\benu}{\begin{enumerate}}
\newcommand{\eenu}{\end{enumerate}}
\newcommand{\bdes}{\begin{description}}
\newcommand{\edes}{\end{description}}
\newcommand{\bt}{\begin{tabular}}
\newcommand{\et}{\end{tabular}}

\newcommand \thetabf{{\mbox{\boldmath$\theta$\unboldmath}}}

\newcommand \gammabf{\mbox{\boldmath$\gamma$\unboldmath}}

\newcommand \zetabf{\mbox{\boldmath$\zeta$\unboldmath}}

\newcommand \bbf{{\bf b}}
\newcommand \cbf{{\bf c}}

\newcommand \fbf{{\bf f}}

\newcommand \hbf{{\bf h}}

\newcommand \lbf{{\bf l}}

\newcommand \tbf{{\bf t}}
\newcommand \ubf{{\bf u}}

\newcommand \ybf{{\bf y}}
\newcommand \zbf{{\bf z}}

\newcommand \Cbf{{\bf C}}

\newcommand \Wbf{{\bf W}}

\newcommand{\Rset}{{\mathbb R}}
\newcommand{\Cset}{{\mathbb C}}

\newcommand{\circlambda}{\mbox{$\Lambda$
             \kern-.85em\raise1.5ex
             \hbox{$\scriptstyle{\circ}$}}\,}

\renewcommand \thetabf{\boldsymbol{\theta}}

\renewcommand \gammabf{\boldsymbol{\gamma}}

\renewcommand \zetabf{\boldsymbol{\zeta}}

\usepackage{color}

\begin{document}

\title{\huge Multi-Band Wi-Fi Sensing with Matched Feature Granularity}

\author{Jianyuan Yu$^{*}$,        Pu Wang$^{*}$,
        Toshiaki~Koike-Akino,
        Ye~Wang, Philip V. Orlik, and R. Michael Buehrer
\thanks{The work of J.~Yu was done during his internship at MERL.}
\thanks{P.~Wang, T.~Koike-Akino, Y.~Wang and P.V.~Orlik are with Mitsubishi Electric Research Laboratories (MERL), Cambridge, MA 02139, USA.}
\thanks{J.~Yu and R.M.~Buehrer are with the Bradley Dept. of Electrical and Computer Engineering, Virginia Tech, Blacksburg, VA 24060, USA.}
\thanks{$^{*}$Equal contribution. Corresponding author: pwang@merl.com}
}

\maketitle

\begin{abstract}
Complementary to the fine-grained channel state information (CSI) and coarse-grained received signal strength indicator (RSSI) measurements, the mid-grained spatial beam attributes (e.g., beam SNR) during the millimeter-wave (mmWave) beam training phase were recently repurposed for Wi-Fi sensing applications such as human activity recognition and indoor localization. This paper proposes a multi-band Wi-Fi sensing framework to fuse features from both CSI at sub-7 GHz bands and the mid-grained beam SNR at 60 GHz with \emph{feature granularity matching} that pairs feature maps from the CSI and beam SNR at different granularity levels with learnable weights. To address the issue of limited labeled training data, we propose to pre-train an autoencoder-based multi-band Wi-Fi fusion network in an unsupervised fashion. For specific sensing tasks, separate sensing heads can be attached to the pre-trained fusion network with fine-tuning. The proposed framework is thoroughly validated by three in-house experimental datasets: 1) pose recognition; 2) occupancy sensing; and 3) indoor localization. Comparison to a list of baseline methods demonstrates the effectiveness of granularity matching. Ablation study is performed as a function of the number of labeled data, latent space dimension, and fine-tuning learning rates.
\end{abstract}

\begin{IEEEkeywords}
Wi-Fi sensing, WLAN sensing, channel state information, beam attributes, fusion, autoencoder, millimeter wave, fingerprinting, deep learning. 
\end{IEEEkeywords}

\section{Introduction}
\label{sec:intro}

Wi-Fi sensing or WLAN sensing has received tremendous attention over the past decade. More recently (September 2020), IEEE $802.11$ Standards Association established a new task group for WLAN sensing for making greater use of $802.11$ technologies towards new industrial and commercial applications in home security, entertainment, energy management (HVAC, light, device power savings), elderly care, and assisted living \cite{11af, XiongJamieson13}. 

Most Wi-Fi sensing frameworks use either fine-grained channel state information (CSI)~\textcolor{black}{\cite{VasishtKumar16, ChenChen17, ChenChen17b, WangGao17, WangGao17b, HsiehChen19, ChenZhang17, WangGao16, XiangZhang19, XiangZhang19b}} or coarse-grained RSSI measurements~\textcolor{black}{\cite{Li06, MazuelasBahillo09, PajovicOrlik15, LiuFang16, BahlPadmanabhan00, KingKopf06, YoussefAgrawala08, BrunatoBattiti05, WuLi07, LiZhang14, HeChan16, KushkiPlataniotis07, FigueraJimenez09, FangLin08, LuZou16, DaiYing16, HoangYuen19}}; see more detailed literature review in the next section. The conventional RSSI measurement suffers from the measurement instability and coarse granularity of the channel information, leading to limited accuracy for localization. The CSI measurement is more fine-grained but requires access to physical-layer interfaces and high computational power to process a large amount of sub-carrier data. 

These limitations motivate a recent adoption of mid-grained intermediate channel measurements which are more informative (e.g., in the spatial domain) than the RSSI measurement and easier to access than the lower-level CSI measurement \cite{PajovicWang19, WangPajovic19, KoikeWang20, WangKoike20b, YuWangKoike20}. Specifically, spatial beam SNRs that are inherently available (with zero overhead) for beam training for IEEE 802.11ad/ay standards operating at millimeter-wave (mmWave) bands \cite{802.11ad, NitscheCordeiro14, GhasempourSilva17}, are used to construct the fingerprinting/training dataset. The use of the mid-grained channel measurement was made possible by earlier efforts in \cite{SteinmetzerWegemer17b, BielsaPalacios18, SahaAssasa18} which enabled easy access to beam SNR measurements from  commercial off-the-shelf (COTS) 802.11ad Wi-Fi routers. 

Fusion-based approaches have been considered before to improve robustness and accuracy. Most fusion-based localization methods make use of heterogeneous sensor modalities such as on-device or wearable sensors, e.g., inertial navigation system (INS), magnetic sensors, accelerometer, ultrasound, and other surrounding sensors such as camera, Bluetooth, and ZigBee. These heterogeneous sensor signals are then fused with Wi-Fi signals (either RSSI or CSI) in the traditional maximum likelihood framework or more advanced deep learning context. For instance, continuous localization of indoor pedestrians was achieved using INS with tracking errors adjusted by the Wi-Fi \cite{ChenWu14}. DeepFusion combines heterogeneous wearable (e.g., smartphone and smartwatch) and wireless (Wi-Fi and acoustic) sensors for human activity recognition \cite{XueJiang19}. A recent survey paper can be found in \cite{GuoAnsari20}. 

When only wireless radio frequency (RF) sensors (e.g., Wi-Fi, Bluetooth, ZigBee, LTE) are considered, the fusion of the coarse-grained RSSI, Bluetooth, and ZigBee measurements was considered in \cite{RodriguesVieira11} with the $k$ nearest neighbor ($k$NN) method. In \cite{AhmedArablouei19}, the fine-grained CSI was used first to extract the Angle-of-Arrival (AoA) measurements and then fused with RSSI from Bluetooth in the context of the traditional maximum-likelihood estimation (MLE) framework. 

If we further narrow down the scope to Wi-Fi-only measurements, the fusion between the CSI and RSSI can be done in a way similar to \cite{AhmedArablouei19} or in a straightforward manner by concatenating the scalar RSSI to the high-dimension CSI. For the CSI Wi-Fi measurements,  \cite{DangSi19} proposed to fuse the phase and amplitude of the fine-grained CSI, as opposed to the magnitude-only CSI, for the purpose of localization. When multiple access points (APs) are available, the fusion of multi-view CSI measurements over APs was considered in \cite{SanamGodrich20} using the generalized inter-view and intra-view discriminant correlation analysis. \cite{GonultasLei20} proposed to fuse the probability maps from multiple APs and multiple transmit antennas. Moreover, \cite{gao2020crisloc} propose the Maximum Mean Discrepancy (MMD) as a transfer learning technique for indoor localization with different placement of the APs.

\begin{figure*}[t]
 \centering
 \subfloat[][Multipath propagation]{\includegraphics[width=0.38\linewidth]{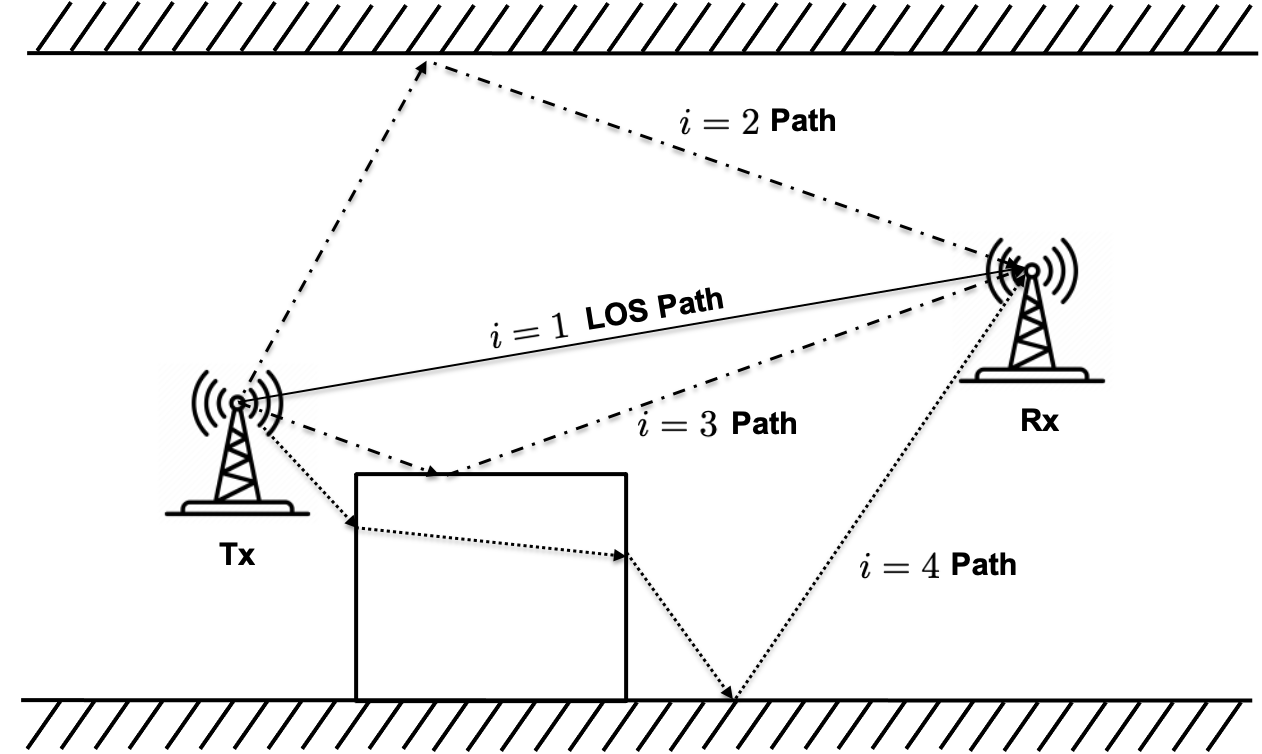}} \ \  
 \subfloat[][Received multipath waveforms]{\includegraphics[width=0.3\linewidth]{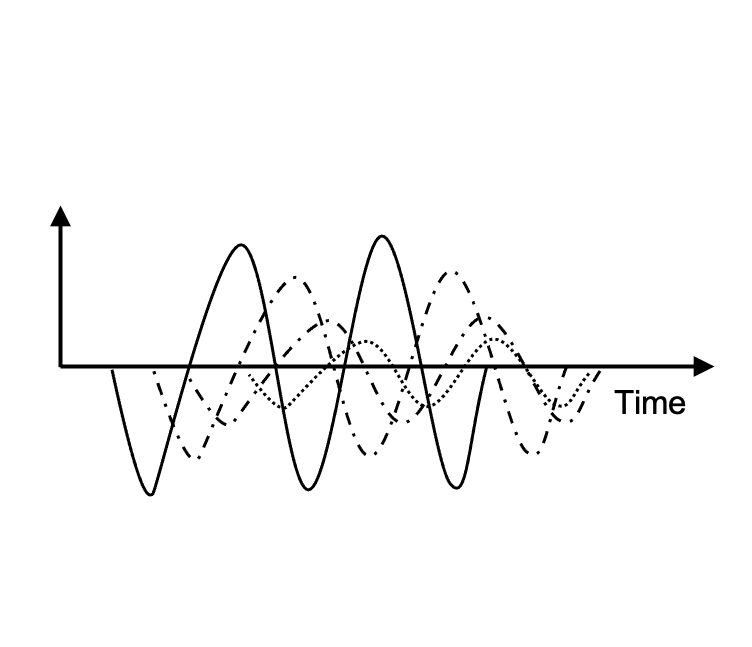}}  \ \ \
 \subfloat[][RSSI and CSI (e.g., CIR \& CFR)]{\includegraphics[width=0.26\linewidth]{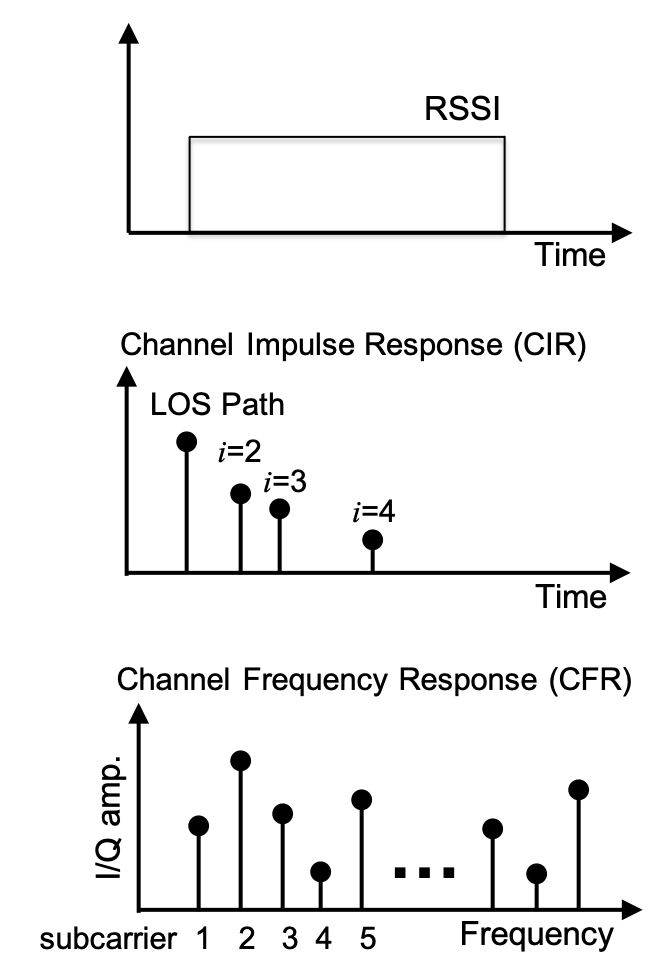}}
\caption{The coarse-grained RSSI and fine-grained CSI channel measurements at sub-$6$ GHz frequency bands, e.g., $2.4$ GHz and $5$ GHz in $802.11$ac.}
\label{fig_measurements}	
\end{figure*}

In this paper, we propose to fuse the fine-grained CSI measurements at sub-6 GHz from multiple spatial streams and the mid-grained beam SNRs at the mmWave band of 60 GHz. Both kinds of Wi-Fi channel measurements provide rich yet complementary features in different physical domains. As complex-valued amplitudes at OFDM subcarrier tones, the CSI measurements are equivalent to the power delay profile (PDP) in the time domain and reflect the power distribution along propagation paths. On the other hand, the beam SNRs are more about spatial-domain channel measurements over various beamforming directions or beamspace. In particular, our contributions and results are summarized as follows:
\begin{itemize}
 \item First work to fuse the two distinct kinds of Wi-Fi channel measurements at different frequency bands. The mid-grained beam SNRs are richer and more stable channel measurements than the coarse-grained RSSI and can lead to more meaningful fusion with the fine-grained CSI. 
    \item Introduce a multi-band Wi-Fi fusion method that accounts for the granularity matching between the CSI and beam SNR measurements. The granularity matching is realized via a learnable fusion block that pairs any two feature maps from the CSI and beam SNR at different feature levels and assigns adaptive weights to the paired feature maps. 
   \item Propose an autoencoder-based multi-band fusion network that can be pre-trained in an unsupervised fashion. With the pre-trained fusion network, we propose attaching multi-task heads to the fused features by fine-tuning the fusion block and training the multi-task heads from the scratch. 
   \item Implement an in-house multi-band Wi-Fi testbed consisting of commercial $802.11$ac- and $802.11$ad-compliant Wi-Fi routers and collect real-world measurements in standard room environments (e.g., apartment and lab)  for three Wi-Fi sensing tasks: 1) pose recognition, 2) occupancy sensing, and 3) indoor localization. 
  \item Conduct comprehensive performance analysis by including four baseline methods and evaluating performance as a function of the number of labeled training data, latent space, and the fine-tuning learning rates. 
\end{itemize}


\section{Wi-Fi Channel Measurements: An Overview}
\label{sec:review}

In the following, we provide a literature review on Wi-Fi channel measurements and their applications in indoor localization, human activity recognition, and wireless sensing. 

%
%

\begin{figure*}[t]
 \centering
 \subfloat[][Multipath propagation]{\includegraphics[width=0.38\linewidth]{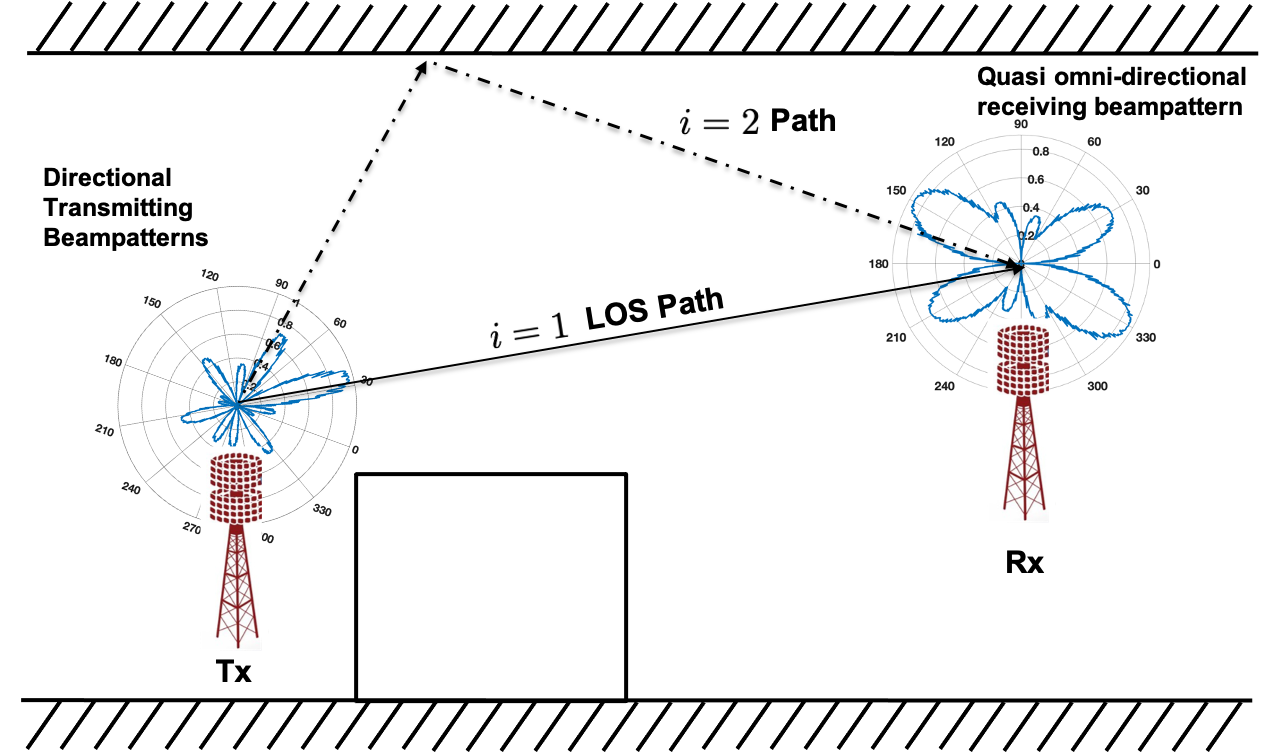}} \ \  
 \subfloat[][Received multipath waveforms]{\includegraphics[width=0.30\linewidth]{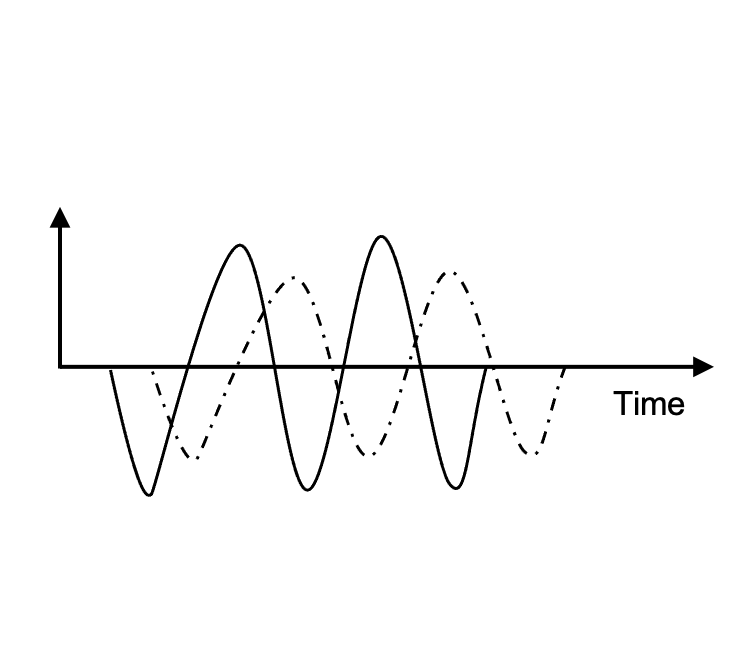}}  
 \subfloat[][Beam SNR]{\includegraphics[width=0.30\linewidth]{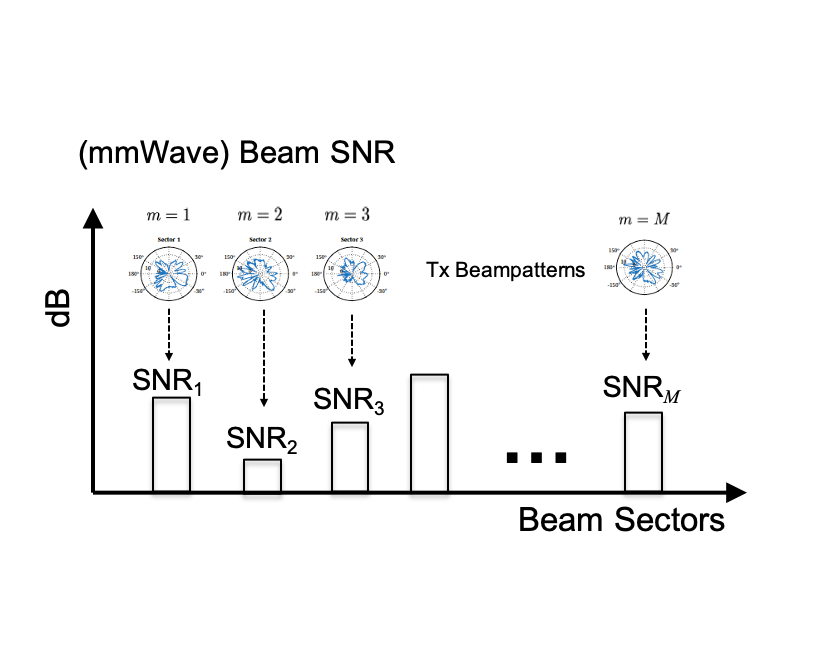}}
\caption{The mid-grained beam SNR measurement at mmWave frequency bands, i.e., $60$ GHz in $802.11$ad. }
\label{fig_bSNR}	
\end{figure*}

\subsection{RSSI: Received Signal Strength Indicator}

In a typical indoor environment, the baseband signal voltage at an RF receiver at a given time is measured as \cite{YangZhou13}
\begin{align} \label{equ_v}
V=\sum_{i=1}^{N}\left\|a_{i}\right\| e^{-j \theta_{i}},
\end{align}
where $a_i$ and $\theta_i$ are the amplitude and phase of the $i$-th multipath component, and $N$ is the total number of multipath components.; see Fig.~\ref{fig_measurements} (a) and (b) for an illustration.  Then RSSI is the received power in decibels (dB) as
\begin{equation} \label{equ_r}
\text{RSSI}=10 \log _{2}\left(\|V\|^{2}\right).
\end{equation}

Early Wi-Fi based indoor localization systems used RSSI measurements to estimate indoor location in a direct localization fashion~\cite{Li06, MazuelasBahillo09, PajovicOrlik15, LiuFang16}. For fingerprinting-based methods, RSSI was used directly as fingerprinting data in systems such as Radar~\cite{BahlPadmanabhan00}, Compass~\cite{KingKopf06}, and Horus~\cite{YoussefAgrawala08} due to easy access to 802.11ac- and 802.11n-compliant devices. Classical machine learning methods, e.g., the $k$-nearest neighbor ($k$NN) and support vector machine (SVM), and probabilistic Bayesian methods have been applied to RSSI fingerprinting measurements \cite{BahlPadmanabhan00, BrunatoBattiti05, WuLi07, KushkiPlataniotis07, YoussefAgrawala08, FangLin08, FigueraJimenez09, LiZhang14, HeChan16}.  Advanced deep learning methods such as multi-layer neural networks \cite{DaiYing16} and recurrent neural networks (RNNs)  \cite{HoangYuen19}  showed improved localization performance over classical machine learning approaches. 

Nevertheless, RSSI measurements, as a simple superposition of multipath voltages \eqref{equ_v}, fluctuate over time at a given location and are only a scalar measurement as shown in \eqref{equ_r}. RSSI is considered a \emph{coarse-grained} channel measurement since it results in a single value per channel.

\subsection{CSI: Channel State Information}

By modeling the wireless channel as a temporal linear filter, one can measure the channel impulse response (CIR) $h(\tau)$ as
\begin{equation}
h(\tau)=\sum_{i=1}^{N} a_{i} e^{-j \theta_{i}} \delta\left(\tau-\tau_{i}\right),
\end{equation}
where $\tau_i$ is the delay of the $i$-th multipath, and $\delta(\cdot)$ is the Dirac delta function. In other words, the channel can be represented by the multipath amplitude and phase distribution at corresponding delays; see the middle figure of Fig.~\ref{fig_measurements} (c) for an illustration. Given the CIR $h(t)$, the received signal $r(t)$ is the temporal convolution of the preamble $s(t)$ and $h(t)$: $r(t) = s(t) \otimes h(t)$. The channel frequency response (CFR) or, equivalently, the Fourier transform of the CIR, is often measured as 
\begin{equation}
H(f)=S^{*}(f) R(f),
\end{equation}
where $R(f)$ is the Fourier transform of $r(t)$ and $S^{*}(f)$ is the complex conjugate of the Fourier transform of $s(t)$. In commercial Wi-Fi devices, a group of sampled CFRs at a list of subcarriers are measured as
\begin{equation}
H(f_k)= \| H(f_k)\| e^{j \angle H(f_k)}, k = 1, 2, \cdots, K,
\end{equation}
where $f_k$ is the $k$-th subcarrier frequency; see the bottom figure of Fig.~\ref{fig_measurements} (c) for an illustration. These complex-valued CFRs are reported to upper network layers in the format of CSI \cite{HalperinHu10}. Compared with the coarse-grained RSSI, the CSI provides a \emph{fine-grained} channel measurement with better capability to resolve multipath in the time or frequency domain. While still suffering from temporal fluctuation, the CSI provides a set of complex-valued random variables (i.e., multipath components) compared with the scalar RSSI. 

Recent efforts on CSI extraction from COTS Wi-Fi devices, e.g., Intel Wi-Fi Link $5300$ radio \cite{HalperinHu11}, Atheros $802.11$n chipsets \cite{XieLi15}, and Cypress $802.11$ac chipsets \cite{GringoliSchulz19}, have enabled access to the CSI over a bandwidth of up to $80$ MHz at sub-$6$ GHz (i.e., $2.4$ and $5$~GHz bands). This has sparked a large swath of learning-based Wi-Fi sensing applications \cite{VasishtKumar16, ChenChen17, ChenChen17b, WangGao17, WangGao17b, HsiehChen19, LuWen16, ZengPathak16, WuZhang17, ZouZhou18, YangZou18,  GuLiu18, GuZhang19, WangJiang19, WangFeng19, WangZhou19, ZhangRuan20}. For instance, ConFi~\cite{ChenZhang17} used convolutional neural networks (CNNs) to train CSI measurements from three antennas, to classify the location and estimate location coordinates. \cite{WangGao16} fingerprinted full CSI over multiple time instants, calibrated their phases,  and fitted one autoencoder for one location. CSI measurements can also be directly trained to regress the coordinate \cite{XiangZhang19, XiangZhang19b}. More recently, \cite{WangFeng19} used annotations from camera images to train fine-grained CSI measurements for pose and human tracking. 


\begin{figure*}[t]
     \begin{center}
	 \includegraphics[width=0.9\textwidth]{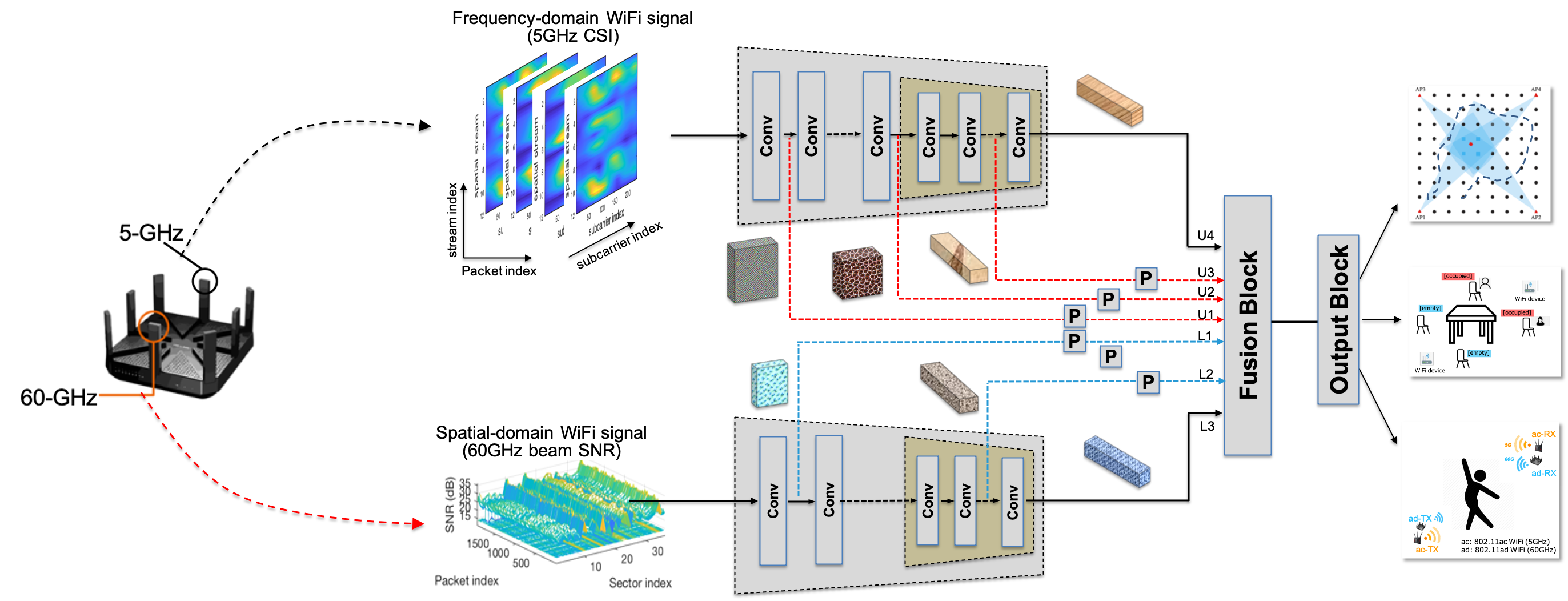} 
	    \end{center}
    \caption{Multi-band Wi-Fi fusion with granularity matching. }
   \label{fig-featurePermutation}
\end{figure*}

\subsection{Beam SNR: mmWave Beam Training Measurement}
At mmWave bands, another type of channel measurements from the mandatory beam training phase provides a \emph{mid-grained} measurement in the beam angle domain for Wi-Fi sensing. During the mandatory beam training phase defined by the $802.11$ ad/ay standards, directional probing beampatterns are used to determine desired directions for subsequent data communication to compensate large path loss at mmWave bands \cite{802.11ad, NitscheCordeiro14, GhasempourSilva17}. For each probing beampattern (also referred to as beam sectors), beam SNR is computed as a measure of beam quality. 

For a given pair of transmitting and receiving beampatterns, corresponding beam SNR can be defined as
\begin{align}
h_m = \mathsf{BeamSNR}_m = \frac{1}{\sigma^2} \sum\limits_{i=1}^N \gamma_m(\theta_i) \zeta_m(\psi_i) P_i,
\end{align}
where $m$ is the index of beampattern, $\theta_i$ and $\psi_i$ are the transmitting and receiving azimuth angles for the $i$th path, respectively,  $P_i$ is the signal power at the $i$th path, $\gamma_m(\theta_i)$ and $\zeta_m(\psi_i)$ 
are the transmitting and receiving beampattern gains at the $i$th path for the $m$th beampattern, respectively, and $\sigma^2$ is the noise variance. Fig.~\ref{fig_bSNR} (a) shows an example of $I=2$ paths between the transmitting side that probes the spatial domain using a directional beampattern and the receiving side which is in a listening mode with a quasi-omni-directional beampattern, while Fig.~\ref{fig_bSNR} (c) shows a number of pre-determined beampatterns used at the transmitting side to compute corresponding beam SNRs. Due to antenna housing and hardware impairments, these beampatterns exhibit fairly irregular shapes at the $60$~GHz band \cite{SteinmetzerWegemer17b, BielsaPalacios18}.

\subsection{CSI versus Beam SNR}
By comparing Fig.~\ref{fig_measurements} (a) with Fig.~\ref{fig_bSNR} (a), several observations can be made between the CSI and beam SNR. First, multipath propagation is richer at sub-$6$ GHz bands (i.e., $4$ paths) than $60$ GHz bands (i.e., $2$ paths). Second, the through-wall path (denoted as dotted lines) in  Fig.~\ref{fig_measurements} (a) does not survive in the $60$ GHz link due to its much shorter wavelength and less capability to penetrate obstacles (e.g., wall). Third, mmWave Wi-Fi devices are equipped with a phased array that enables highly directional beampatterns while antenna elements of sub-$6$ GHz Wi-Fi devices are mostly in an omni-directional mode. Although multiple antenna elements at sub-$6$ GHz can be used for beamforming, the beampatterns are less directional than the mmWave Wi-Fi device due to the smaller number of antennas and the relatively large inter-element spacing. Consequently, these two distinct kinds of Wi-Fi channel measurements probe the scene with different lenses and may exhibit complementary channel features in different physical space (time/frequency versus space) and domains (omni-directional antennas versus directional beampatterns).

\section{Multi-Band Wi-Fi Sensing with Supervised Learning}
\label{sec:algorithm}
In this section, we introduce the multi-band Wi-Fi fusion framework that combines the features from both the fine-grained CSI measurements and the mid-grained beam SNRs. Particularly, we account for the feature granularity by designing a learnable granularity matching network that pairs multi-scale features from the CSI and beam SNR and assigns linear weights to the paired features. Specifically, Fig.~\ref{fig-featurePermutation} shows the multi-band Wi-Fi sensing network that consists of three blocks: 1) feature extraction, 2) feature fusion with granularity matching, and 3) separate output blocks for individual sensing tasks. Such a framework is then trained end-to-end for human activity recognition, occupancy detection, and localization tasks, respectively, in a supervised fashion. In the following, we introduce each block in more detail.

\subsection{Preliminary}
Like traditional fingerprinting-based Wi-Fi sensing methods, we follow the standard procedure by collecting both CSI and beam SNR measurements corresponding to each class (e.g., pose, occupancy pattern, and location/orientation) as the fingerprinting data. Specifically, we use $\mathbf{C} \in \Cset^{N_s \times M_s}$ and $\mathbf{h} \in \Rset^{M \times 1}$ to denote the CSI measurements from $N_s$ spatial streams over $M_s$ subcarriers and the beam SNRs from $M$ beampatterns. For a given class $l$, $R$ fingerprinting snapshots, ${\mathbf{C}}_1(l), \ldots, {\mathbf{C}}_R(l)$ and ${\mathbf{h}}_1(l), \ldots, {\mathbf{h}}_R(l)$, are collected to construct the offline training dataset. By collecting many realizations of both Wi-Fi channel measurements at $L$ classes, we will have $L$ sets of training data in the training dataset. For the labeled data, the label class $l$ is automatically attached to both $\Cbf_r(l)$ and $\hbf_r(l)$.  

%
%
%

\subsection{Feature Extraction}
In Fig.~\ref{fig-featurePermutation}, the feature extraction block employs two separate convolution networks to encode the feature maps for CSI-only and beam SNR-only measurements. Specifically, 
\begin{align} \label{ff}
{\ybf}^c_{\ell} & = f^c_{\ell}({\ybf}^c_{\ell-1}, {\thetabf}^c_{\ell}), \quad \ell = 1, 2, \ldots, N_c  \notag \\
{\ybf}^h_{\ell} & = f^h_{\ell}({\ybf}^h_{\ell-1}, {\thetabf}^h_{\ell}), \quad \ell = 1, 2, \ldots, N_h, 
\end{align}
where $\ybf^c_0=\Cbf_r(l)$, $\ybf^h_0=\hbf_r(l)$, and $f^{c/h}_{\ell}$ denotes the convolution operation (including batch normalization, pooling and activation functions) for CSI-only and beam SNR-only feature extraction with kernel parameters $\thetabf^{c/h}_{\ell}$ at the $\ell$-th layer. Overall, we use $N_c$  and $N_h$ convolution layers for, respectively, the CSI and beam SNRs to gradually shrink the dimension of the feature maps while increasing the number of feature maps. The $\ell$-th feature map is denoted as $\mathbf{y}^{c/h}_{l} \in \Rset^{N^{c/h}_{\ell} \times H^{c/h}_{\ell} \times W^{c/h}_{\ell}}$, where $\{ N^{c/h}_{\ell},  H^{c/h}_{\ell}, W^{c/h}_{\ell}\}$ are, respectively, the number of channels, height and width of the $\ell$-th feature map. Note that the number of layers for CSI can be different from that for beam SNRs, i.e., $N_c \neq N_h$. 

\begin{figure}[t]
     \begin{center}
	 \includegraphics[width=0.45\textwidth]{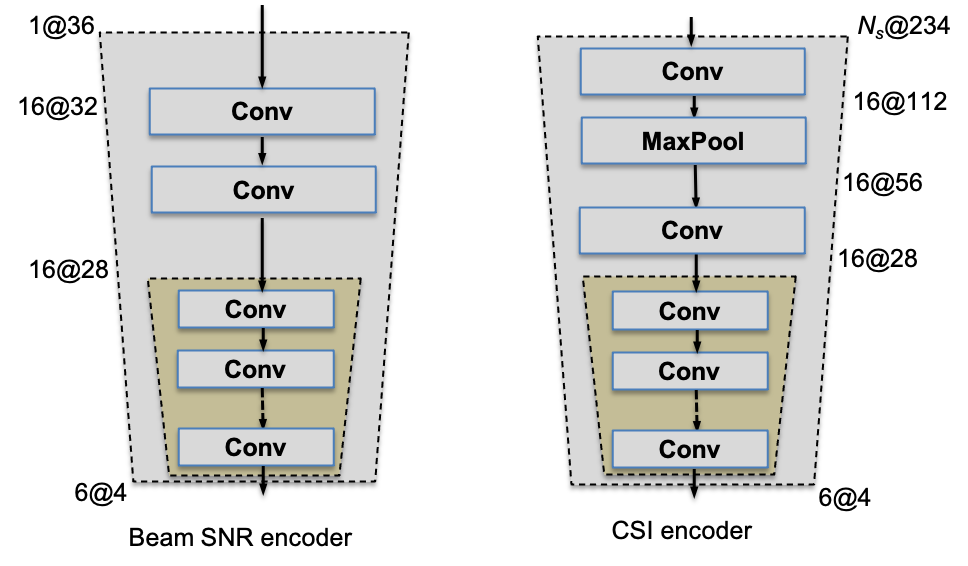} 
	    \end{center}
    \caption{Separate CSI and beam SNR encoder architectures. }
   \label{fig-encoder}
\end{figure}

To be more specific, we use $5$ convolutional blocks, respectively, for the CSI and beam SNR encoders, as shown in Fig.~\ref{fig-encoder}.  Each convolution block consists of a standard convolution layer followed by a batch normalization layer and a rectified linear unit (ReLU) activation layer. For the CSI encoder, we use the first $2$ convolution blocks and $1$ maxpooling layers to increase the number of feature maps to $16$ and downsize the dimension to $28$. Similarly, the beam SNR encoder uses $2$ convolution blocks to downsize the input to the same feature space. Then the $3$ subsequent convolution blocks share the same structure as highlighted in tan colored blocks of  Fig.~\ref{fig-encoder}.

\subsection{Feature Fusion}
Since the CSI and beam SNRs are two distinct Wi-Fi channel measurements with different channel feature granularities, one needs to take into account the granularity correspondence for the multi-band Wi-Fi fusion.  For instance, CSI features at later convolution layers may have a better correspondence to beam SNR features at earlier layers as the CSI is likely to have finer granularity than the beam SNR. To this end, we extract multi-scale feature maps at selected convolution layers as two sets of threads
\begin{align}
    \{\ybf^c_{\ell \in \mathbb{L}_c}\}\quad \& \quad \{\ybf^h_{\ell \in \mathbb{L}_h}\},
\end{align} 
where $\mathbb{L}_c$ and $\mathbb{L}_h$ denote the selected layer indices of the CSI and, respectively, beam SNR encoders.  For instance of Fig.~\ref{fig-featurePermutation}, we have $4$ layer indices in $\mathbb{L}_c = \{1, 3, 5, 6\}$ and $3$ elements in  $\mathbb{L}_h = \{1, 4, 5\}$. 

\begin{figure}[t]
     \begin{center}
	 \includegraphics[width=0.32\textwidth]{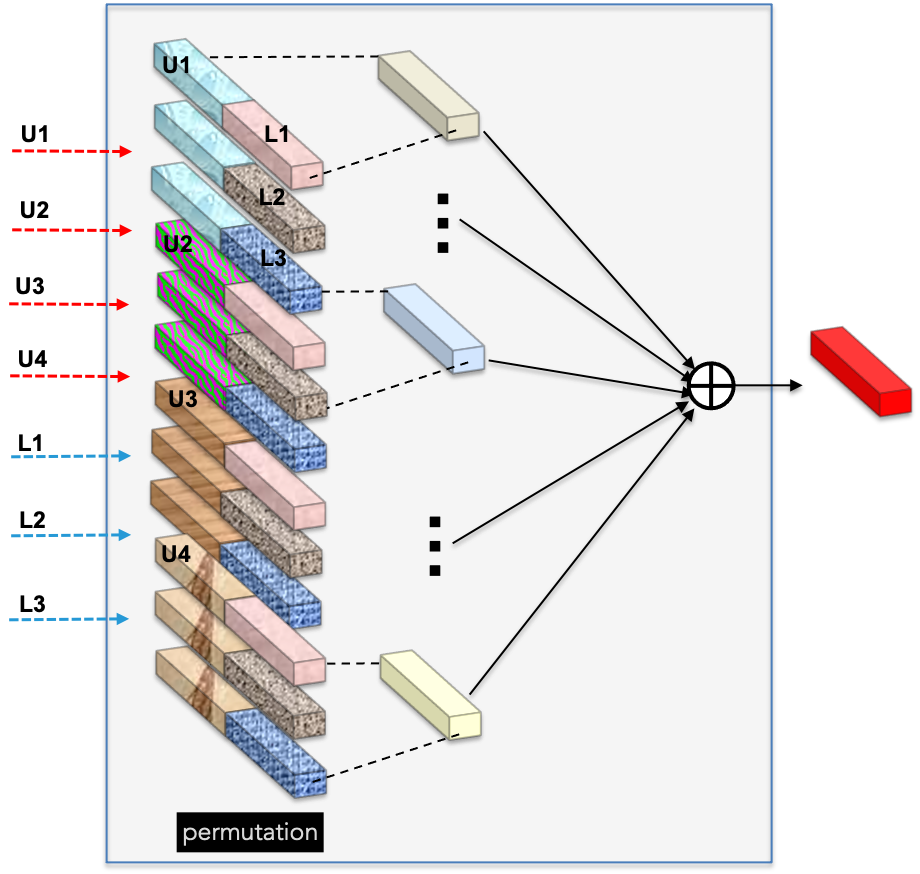} 
	    \end{center}
    \caption{Feature fusion with granularity matching. }
   \label{fig-permutation}
\end{figure}

With an identity mapping or aa linear projection to a smaller dimension, the features from the upper encoder branch are mapped to $\{\ubf_\ell\}_{\ell=1}^{|\mathbb{L}_c|}$ and the lower ones as  $\{\lbf_\ell\}_{\ell=1}^{|\mathbb{L}_h|}$ in Fig.~\ref{fig-featurePermutation}, where $| \mathbb{L} |$ denoting the cardinality of the set $\mathbb{L}$. To find correspondences between the extracted multi-scale features from the CSI and beam SNR, the proposed fusion scheme concatenates two selected feature maps from the two encoders: one is from the CSI encoder and the other from the beam SNR encoder.
\begin{align}
\ybf_p = \left [\ubf_{\ell};  \lbf_{\ell'} \right], \ell \in \{1, \cdots, | \mathbb{L}_c | \}, \ell' \in \{1, \cdots, | \mathbb{L}_h | \},
\end{align}
where $p=1, \cdots, P$ denotes the fusion pair index with $P=| \mathbb{L}_c | |\mathbb{L}_h |$. For instance, we have $P=12$ pairs in Fig.~\ref{fig-permutation} by pairing one of four CSI features $\ubf_{l}$ and one of three beam SNR features $\lbf_{\ell'}$. 

Then, each concatenated feature is fused with a fully connected layer into the same latent dimension $d$, 
\begin{align} 
\fbf_p = \Wbf_p \ybf_{p} + \bbf_p, \quad p = 1, \cdots, P,
\end{align}
where $\Wbf_p$ is the projection weight matrix for the $p$-th pair with $\bbf_p$ denoting the bias term. The final fusion layer is a linear combination of the fused feature map as
\begin{align} \label{lw}
\fbf =  \sum\limits_{p=1}^P a_p \fbf_p, 
\end{align}
where $a_p$ is the fusion weight that can be learned from the training data. The fusion weights acts as a soft selection vector on all concatenated feature vectors. A higher weight implies that the corresponding pair of features has a better granularity matching level. In the case of $\{a_p\}_{p<P}=0$ and $a_P=1$, the fused feature map \eqref{lw} only combines the features at the final layers of the two encoders. In contrast, with non-zero $\{a_p\}_{p<P}$, the proposed fusion scheme allows multi-scale features from earlier layers to find its correspondence from the other branch and contribute to the final fused features.  

\subsection{Output Block}
In Fig.~\ref{fig-featurePermutation}, the fused feature map $\fbf$ is fed into the output block to generate the output vector ${\ubf}$. The output block consists of a number of fully-connected layers:
\begin{align} \label{ob}
{\zbf}_{\ell}= g_{\ell}(\mathbf{z}_{\ell-1}, {\gammabf}_{\ell}), \quad \ell = 1, 2, \ldots, N_g,
\end{align}
where $\zbf_0 = \fbf$,  $\ubf={\zbf}_{N_g}$, and $g_{\ell} = \mathbf{\sigma}_{\ell}(\Wbf_{\ell} \zbf_{\ell} + \bbf_{\ell})$ with $\gammabf_\ell = \{\Wbf_\ell, \bbf_{\ell} \}$ and $\mathbf{\sigma}_{\ell}$ denoting the activation function such as ReLU for $\ell < N_g$ and an identity mapping when $\ell =N_g$ at the final output layer. 

\subsection{Cost Function}
To train the multi-band Wi-Fi sensing network with labeled training data
$\{\Cbf_r(l), \hbf_r(l) \}$ with the label $l$, corresponding output of the last layer $\mathbf{u}$ is first normalized with the softmax \textcolor{black}{operation} as
\color{black}
\begin{align}
{{s}}_n = {\exp(u_n)} \big/ { \sum\limits_{i=1}^N \exp(u_i)}, \quad n \in \{1, 2, \ldots, N\}. 
\end{align}
\color{black}
where $s_n$ is the $n$th element of the normalized output $u_n$. Then, the cross-entropy loss function is computed over the score vector ${\mathbf{s}} = [s_1, s_2, \ldots, s_N]$ and the corresponding one-hot label vector $\mathbf{c} = [c_1, c_2, \ldots, c_N]$ as
\begin{align} \label{claCF}
L_{\text{classification}} = - \sum\limits_n c_n \log(s_n).
\end{align}
where the one-hot label vector $\cbf$ is $1$ at the $l$-th element and $0$ elsewhere. 
Note that, for different sensing applications, the dimension of the output vector $\ubf$ may be different, depending on the number of classes. For instance, $N=8$ for the pose recognition and occupancy sensing and $N=16$ for the indoor localization.


\begin{figure*}[t]
     \begin{center}
	 \includegraphics[width=0.92\textwidth]{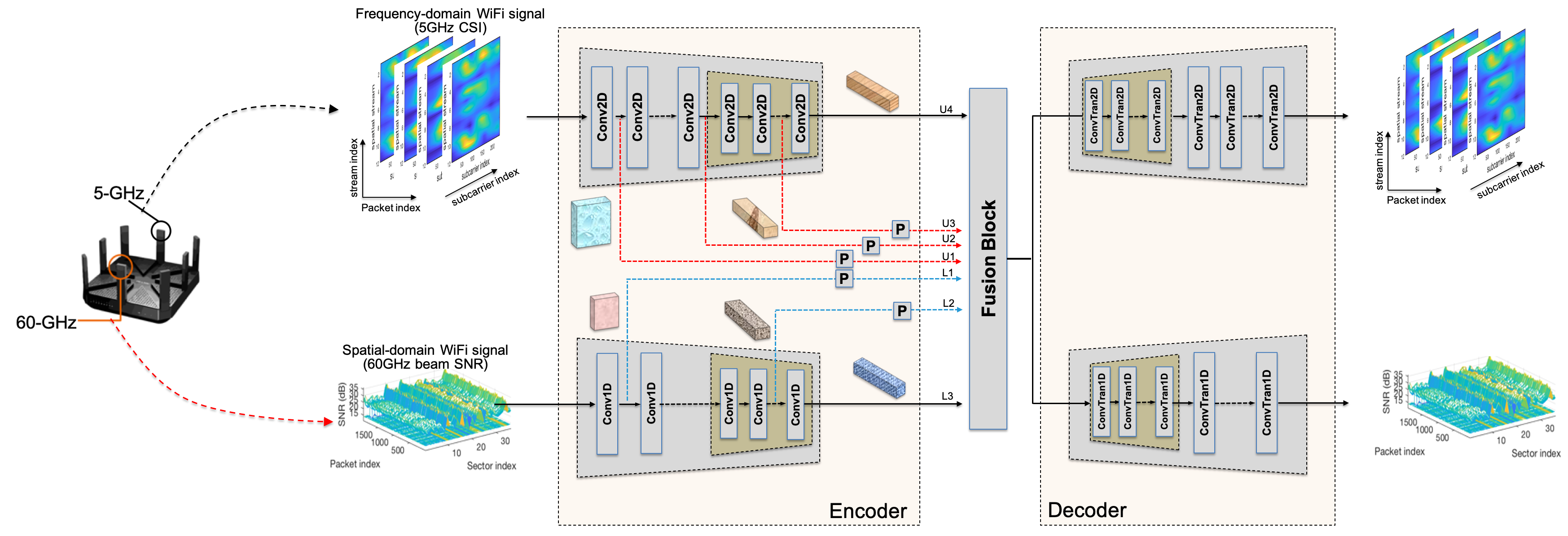} 
	    \end{center}
    \caption{Unsupervised training of feature extraction and fusion blocks with unlabeled (``listening") multi-band Wi-Fi data pooled from multiple sensing tasks. }
   \label{fig-fusion}
\end{figure*}

\section{Multi-Band Wi-Fi Sensing with Transfer Learning}

Although it is simple to train the above network separately for each sensing task, the offline fingerprinting phase is time- and manpower-consuming \cite{ZhaoHan18, JiangLiu19}. To label the data, one has to associate both channel measurements with the ground-truth labels, in the form of pose gesture, occupancy pattern, or user location. This issue is severe when one needs to build up the training dataset when the number of classes is large. Even worse, after a Wi-Fi sensing system is deployed, the ground-truth information is hard to obtain, while the Wi-Fi devices can still gather or ``listen" the environment and access to multi-band Wi-Fi channel measurements. 

To meet the challenges of limited labeled training data, we consider an unsupervised fusion approach using a multi-stream autoencoder to train the encoder and fusion blocks with unlabeled training data $\Cbf_{r}$ and $\hbf_r$ with $r > R$, possibly pooled from multiple sensing tasks. With the unsupervised trained encoder and fusion blocks,  a transfer learning approach is then applied to fine-tune the fusion block and retrain a new attached output block for individual sensing task with limited labeled data $\Cbf_{r}(l)$ and $\hbf_r(l), r=1, \cdots, R$.   

\subsection{Unsupervised Multi-Stream Autoencoder with Feature Granularity Matching} 
Fig.~\ref{fig-fusion} shows a multi-stream autoencoder architecture that uses the same two-branch encoder as in the previous section to extract the features of the two Wi-Fi channel measurements, the same fusion block taking multi-scale features from the two encoder branches and matching the feature granularity, and a new two-branch decoder to synthesize the two input Wi-Fi channel measurements from the same fused feature at the output of the fusion block. 

The multi-stream autoencoder can be trained end-to-end using unlabeled (``listening") multi-band Wi-Fi data that can be pooled from multiple sensing tasks. The motivation here is that, without any class label, one can still train or update the feature extractor and fusion blocks continuously at local Wi-Fi devices or edge units. Whenever new labeled data are available (e.g., users share their location sporadically), the limited amount of labeled data can be used to train the output block for the task-specific sensing task while leaving the training of feature extraction and fusion blocks to the unlabeled data. In the following, we elaborate on each building block of Fig.~\ref{fig-fusion}. 

\subsubsection{Encoder} 
At the encoder side, we re-use the same two encoder networks to extract the feature maps for CSI and beam SNR measurements in Fig.~\ref{fig-encoder}. Multi-scale features from multiple levels of both encoder branches are extracted as $\ybf = \{\ybf^c_{\ell \in \mathbb{L}_c},  \ybf^h_{\ell \in \mathbb{L}_h}\}$ for the feature fusion with granularity matching. 

\subsubsection{Fusion} 
For the fusion block, we also re-use the feature fusion in \eqref{lw} that takes the selected feature maps $\ybf_p$ to generate the fused feature map $\fbf$ of \eqref{lw} with the granularity matching scheme in Fig.~\ref{fig-permutation}. 

\subsubsection{Decoder} 
At the decoder side, the fused feature $\fbf$ is fed to two separate decoders to recover the input CSI and beam SNR measurements simultaneously. Particularly, 
\begin{align} \label{dec}
\fbf & = {\tbf}^c_{0} = {\tbf}^h_{0}, \notag \\
{\tbf}^c_{\ell} & = d^c_{\ell}({\tbf}^c_{\ell-1}, {\zetabf}^c_{\ell}), \quad \ell = 1, 2, \ldots, N_c  \notag \\
{\tbf}^h_{\ell} & = d^h_{\ell}({\tbf}^h_{\ell-1}, {\zetabf}^h_{\ell}), \quad \ell = 1, 2, \ldots, N_h, 
\end{align}
where $\fbf$ is the output of the fusion block in Fig.~\ref{fig-fusion} that is simultaneously fed into the two decoders, $d^{c/h}_{\ell}$ denotes the transposed convolution or upsampling operation at the $\ell$-th layer of the two decoders with corresponding parameters $\zetabf^{c/h}_{\ell}$. Usually, we implement the decode network with a mirrored architecture of its encoder network. Therefore, we keep the same number of layers, $N_c$ and $N_h$, in the decoders. The decoder architectures for the beam SNR and CSI mirror their corresponding encoder architectures in Fig.~\ref{fig-encoder}. 

\subsubsection{Cost Function}
To train the autoencoder-based fusion network with unlabeled data, we adopt a weighted mean-squared error (MSE) as the cost function:
\begin{align} \label{wmse}
L_{\text{MSE}}(w_c) = \lambda \sum\limits_r \left(\ybf^c_0 - \tbf^c_{N_c} \right)^2 +   (1-\lambda) \sum\limits_r \left(\ybf^h_0 - \tbf^h_{N_h} \right)^2
\end{align}
where $\ybf^c_0=\Cbf_r$ and $\ybf^h_0=\hbf_r$ are the CSI and beam SNR training samples, $ \tbf^{c/h}_{N_{c/h}}$ are corresponding outputs of the two decoders, and $\lambda$ is the hyperparameter to balance the reconstruction error between the CSI and beam SNR branches. 

\subsection{Transfer Learning with Limited Labeled Data} 
\begin{figure*}[t]
     \begin{center}
	 \includegraphics[width=0.92\textwidth]{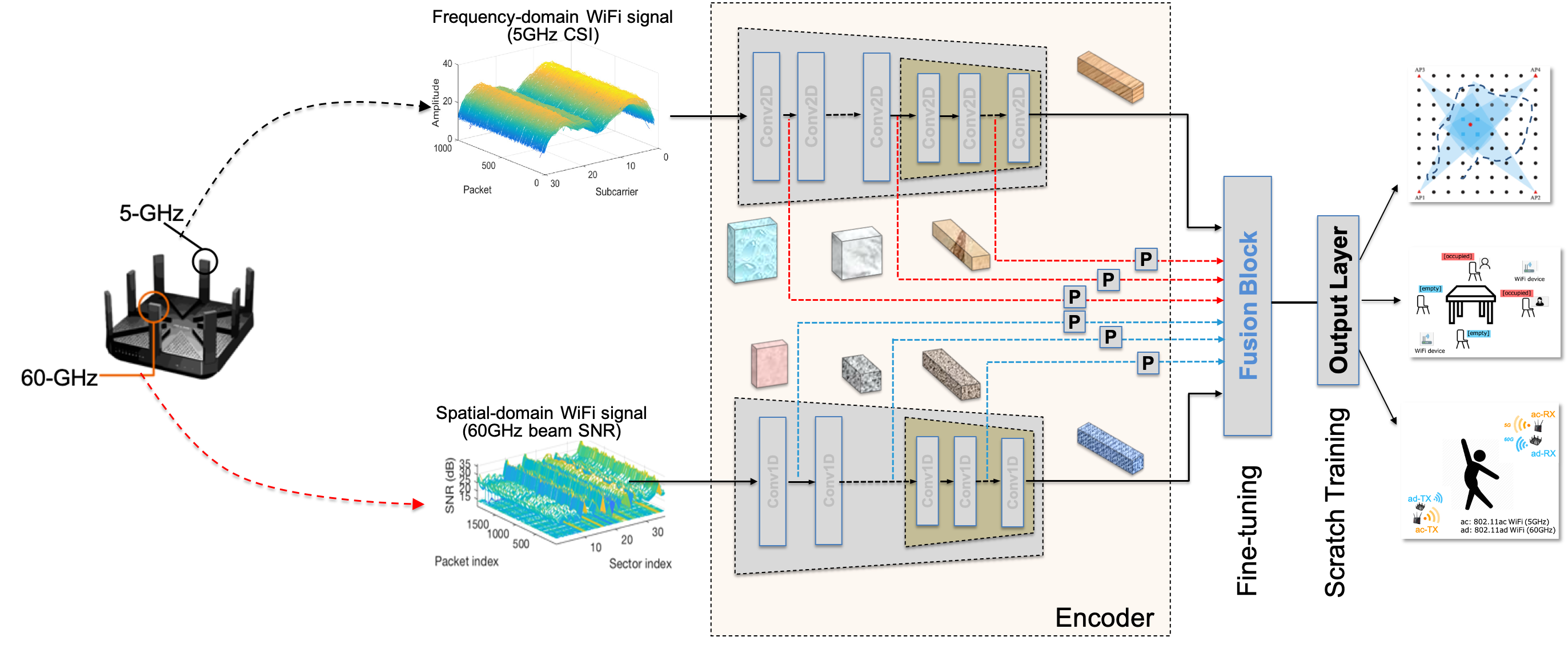} 
	    \end{center}
    \caption{Multi-band Wi-Fi transfer learning with the pretrained encoder from Fig.~\ref{fig-fusion}, fine-tuned fusion block from Fig.~\ref{fig-fusion}, and newly added task-specific output block trained using limited labeled data. }
   \label{fig-TL}
\end{figure*}

Once the autoencoder-based multi-band Wi-Fi fusion network in Fig.~\ref{fig-fusion} is trained, we use the transfer learning to freeze the encoder parameters, fine-tune the fusion block, remove the decoder block, and attach a new output classification block for each sensing task, as shown in Fig.~\ref{fig-TL}. 

With the labeled training data for each sensing task, we retrain the output block which takes the same form of \eqref{ob} and outputs the vector $\ubf$. Specifically, the transfer learning computes the  classification cost function of \eqref{claCF} and back-propagates the gradient of the cost function with respect to the parameters in the output block in a large step size, e.g., a learning rate of $0.01$, while fine-tuning the parameters within the fusion block with a small training step size (e.g., a learning rate of $0.001$ or smaller) on the linear fusion weights $\{ a_p \}_{p=1}^P$ of \eqref{lw} and an even smaller learning rate on $\{\Wbf_p, \bbf_p\}$ of \eqref{ff}. It is worth noting that, for each sensing task, the transfer learning needs to apply separately to train separate output blocks. Once the entire network of Fig.~\ref{fig-TL} is re-trained, we deploy the network with test datasets for different sensing tasks.

\section{Performance Evaluation}
\label{sec:performance}
In the following, we present the performance evaluation for the proposed multi-band Wi-Fi fusion using real-world experimental datasets for three sensing tasks: 1) pose recognition; 2) occupancy sensing; and 3) indoor localization. For each of these three tasks, we compare the proposed fusion methods with the CSI-only and beam SNR-only methods.

\subsection{Experiment Setup}

\begin{figure*}
\centering
 \subfloat[][Pose recognition]{\includegraphics[width=0.32\linewidth]{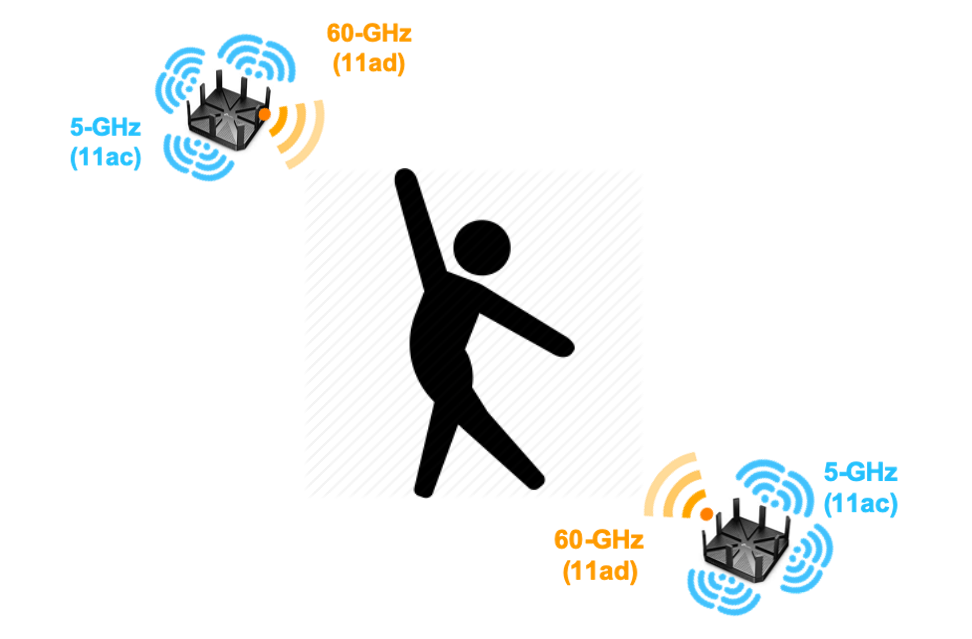}} 
 \subfloat[][Occupancy sensing]{\includegraphics[width=0.38\linewidth]{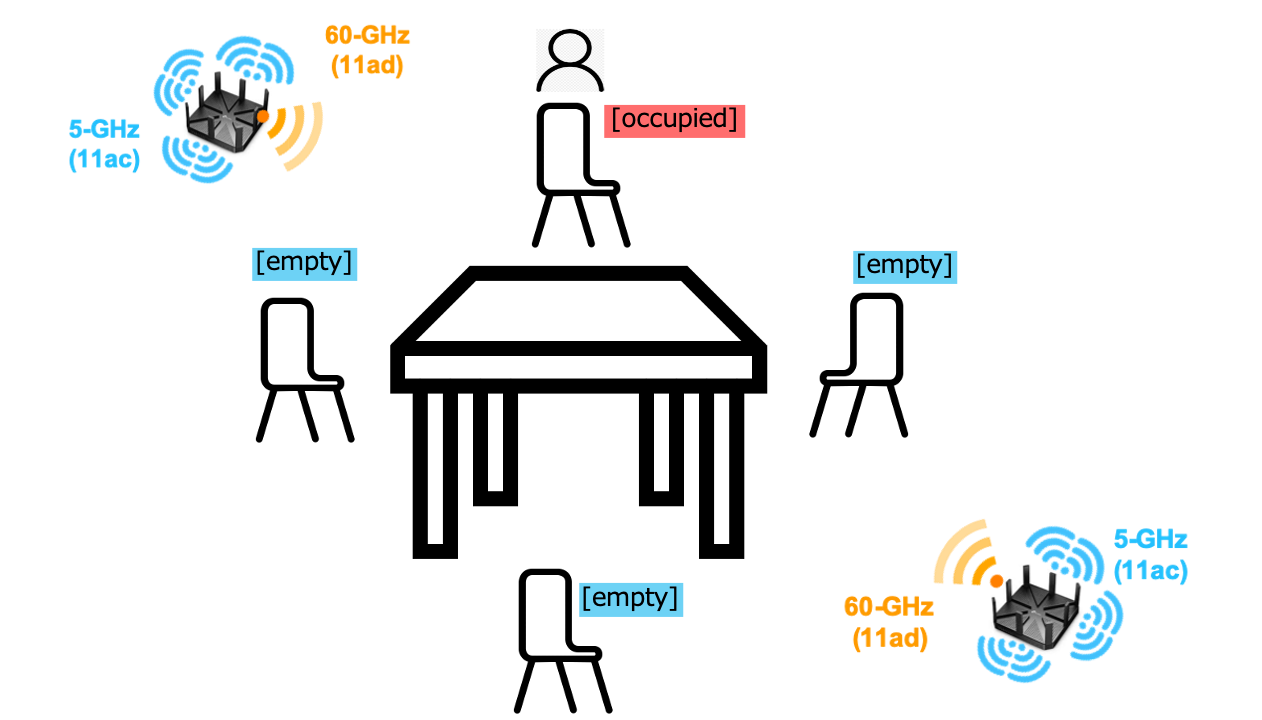}}
  \subfloat[][Indoor localization]{\includegraphics[width=0.33\linewidth]{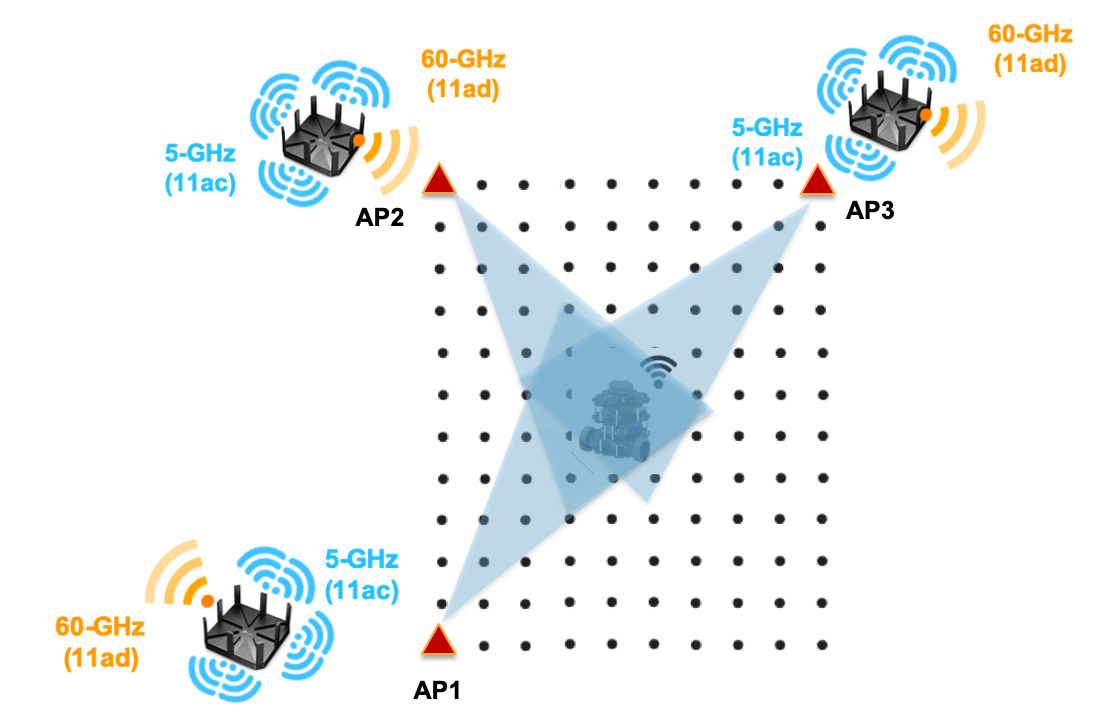}}\\
\subfloat[][Pose snapshots]{\includegraphics[width=0.25\linewidth]{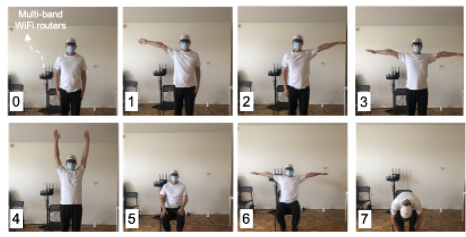}} \ \ \ \ \ \
\subfloat[][Occupancy patterns]{\includegraphics[width=0.35\linewidth]{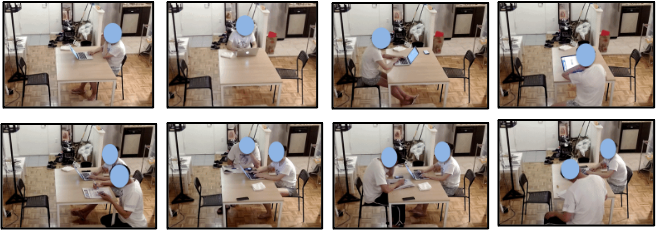}}   \ \ \ \ \ \ \ \ \
 \subfloat[][Fingerprinting grids]{\includegraphics[width=0.18\linewidth]{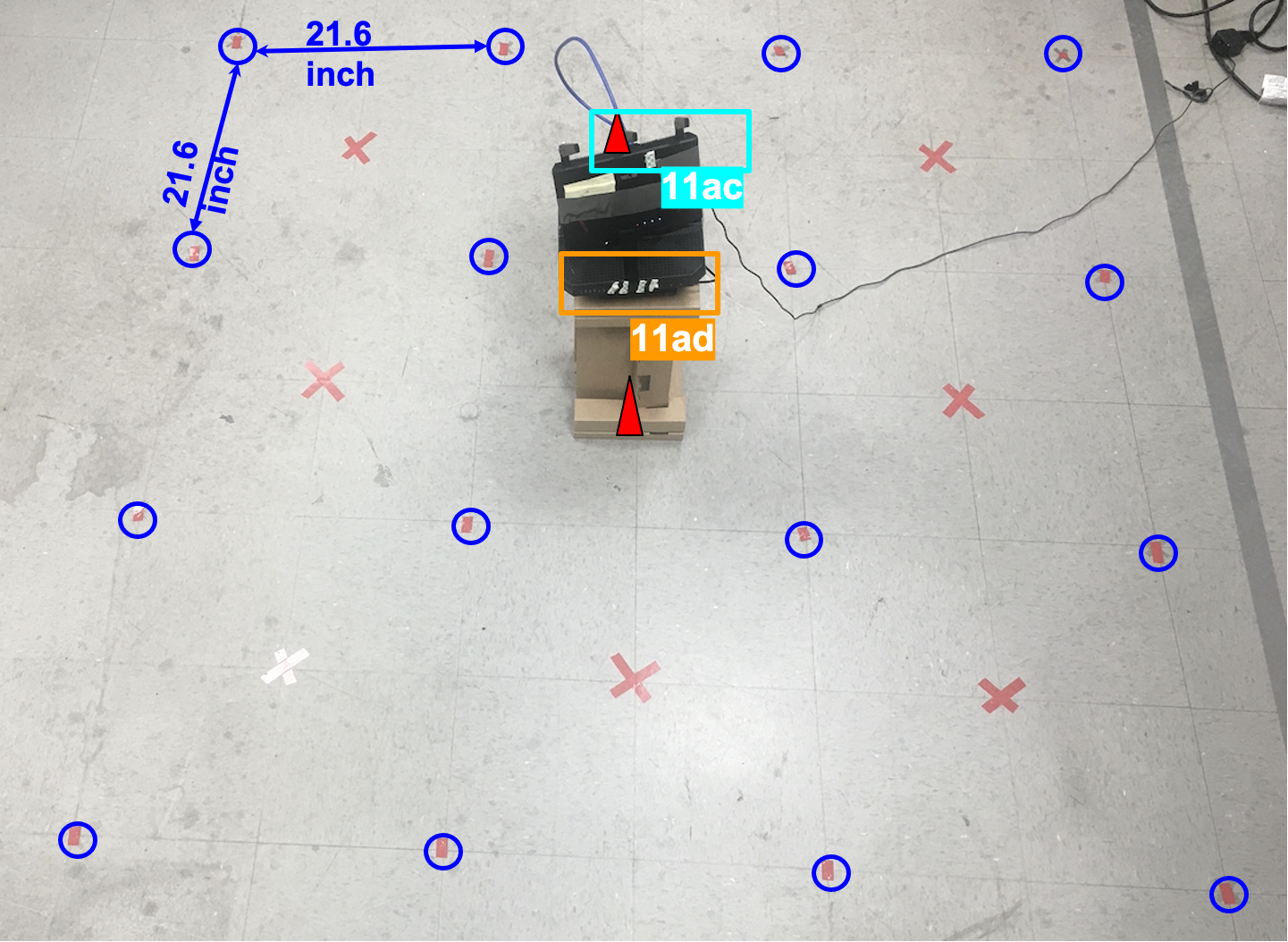}} 
\caption{Multi-band Wi-Fi sensing tasks and experiment settings.  }
\label{fig-tasks}
\end{figure*}

The data collection system consists of multiple commercial $802.11$ac- and $802.11$ad-compliant routers and devices in a configuration shown in Fig.~\ref{fig-tasks} (a)-(c) for the three sensing tasks. Both 5 GHz CSI and 60 GHz beam SNR measurements are recorded in the routers and sent to a workstation via Ethernet cables. The data collection system is deployed in standard indoor room settings, as shown in Fig.~\ref{fig-tasks} (d)-(f).

\textbf{CSI from 802.11ac devices}: We use ASUS RT-AC86U routers with $3$ external and $1$ internal antennas to extract the CSI measurements at $5$ GHz and modified its firmware using the Nexmon CSI Extractor Tool of \cite{GringoliSchulz19}. It allows per-frame CSI extraction for up to $4$ spatial streams using all four receive chains on Broadcom and Cypress Wi-Fi chips with up to 80 MHz bandwidth in both the 2.4 and 5 GHz bands. It also supports devices such as the low-cost Raspberry Pi platform and mobile platforms such as Nexus smartphones. Other CSI tools are also available at $2.4$ GHz and $5$ GHz \cite{HalperinHu11, XieLi15}, but none of them supports $80$ MHz wide channels within the newer $802.11$ac standard (with respect to $802.11$n/g). In addition, it supports various Tx-Rx antenna configurations, up to $4 \times 4$ MIMO. In our in-house testbed, we use a pair of ASUS routers to record all $16$ spatial streams over $242$ subcarriers. However, to mimic a standard mobile terminal, we only use $3$ spatial streams which are equivalent to a configuration of using $1$ ASUS routers with $3$ external antennas and $1$ mobile user with a single antenna.

\textbf{Beam SNR from 802.11ad devices}: We use TP-Link Talon AD7200 routers to collect beam SNRs at $60$ GHz.  Complying with IEEE 802.11ad standards, this router implements Qualcomm QCA9500 transceiver that supports a single stream communication in $60$~GHz range using analog beamforming over $32$-element planar array. The raw beam SNRs are extracted by using the open-source software package in~\cite{SteinmetzerWegemer18}. By matching the patterns of IEEE $802.11$ad beam training frames with the memory inside the chip, one can identify parts of the firmware handling the beam training frames and extract beam SNR measurements from these memory addresses. For Talon AD7200 routers, the beam SNR measurements are further quantized in a stepsize of $0.25$~dB. Overall, from one beam training, one AP can collect $M=36$ beam SNRs for $36$ transmitting beampatterns. It is noted that the resulting beampatterns depart from the theoretical ones and exhibit fairly irregular shapes due to hardware imperfections and housing at $60$~GHz.

\subsection{Sensing Tasks and Datasets}
With the above multi-band Wi-Fi testbed, we conducted multiple sensing tasks as follows:

\subsubsection{Pose Recognition}
Fig.~\ref{fig-tasks} (a) shows the experimental configuration for the pose recognition by using one Wi-Fi station (including both $802.11$ac and $802.11$ad routers) in the front of the subject and one station behind the subject. Both stations are placed on a stand of a height of $1.20$~meters with a distance of approximately $2$ meters. As shown in Fig.~\ref{fig-tasks} (d), the subject is asked to perform a total of $8$ poses including distinct gestures like `sit', `stand with left arm lifted', `stand with both arms lifted', etc. For each pose, we recorded $7$ independent sessions with different time durations and with sufficient time separation in between consecutive two sessions. We grouped the measurements in the first four data sessions as the training data and those in the last three sessions as the test data for the maximum time separation. The numbers of both CSI and beam SNR samples (in brackets) in the training and test datasets are listed in the second and third columns of Table~\ref{table_sampleNumber}. 

\subsubsection{Occupancy Sensing}
Fig.~\ref{fig-tasks} (b) shows the experimental configuration for occupancy sensing around a table in a small living room environment. Two stations are placed in the top left and bottom right corners in a diagonal configuration. The selected $8$ occupancy patterns cover both a single subject sitting on one of the four chairs and two subjects, as shown in Fig.~\ref{fig-tasks} (e). During the data collection, the subject is free to move body parts such as typing the laptop and reading the paper as long as the subject is fixed at the location. We also recorded $7$ independent sessions with sufficient time separation in between consecutive two sessions and grouped the measurements as the training and test datasets in the same way as the pose recognition task. The numbers of both CSI and beam SNR samples (in brackets) in the training and test datasets are listed in the fourth and fifth columns of Table~\ref{table_sampleNumber}. Similar data augmentation is performed to align the beam SNR samples with the higher density CSI data in the training dataset. 

\subsubsection{Indoor Localization}

Finally, we consider indoor localization in a standard lab environment with multiple AP stations (including one $802.11$ac and three $802.11$ad routers) and one user on the grid, as denoted by blue dots in Fig.~\ref{fig-tasks} (f). The AP stations are placed around the three corners of the fingerprinting area, denoted as triangles in Fig.~\ref{fig-tasks} (c). The number of grids is $4 \times 4 =16$ with a grid size of $~54.86$ cm. For each grid, we place another station with both  $802.11$ac and $802.11$ad devices acting as a single-antenna mobile user and record multiple data sessions for training and independent testing. 


\subsection{Wi-Fi Signal Preprocessing}
For the fingerprinting dataset, we perform preprocessing steps to calibrate CSI and beam SNR measurements. 
For the CSI measurements from commercial routers (ASUS RT-AC$86$U in our case), known hardware issues include \cite{XieLi15, KotaruJoshi15, ZhuoZhu17, ZhuZhuo18, ZhangHu20}
\begin{enumerate}
    \item Power control uncertainty: The raw CSI amplitude is compensated by the automatic gain controller (AGC) and mixed with a power amplifier uncertainty error, causing an amplitude offset;
    \item Sample frequency offset (SFO): the sampling frequencies between the transmitter and receiver also exhibit an offset due to non-synchronized clocks, causing a time shift to the digitally sampled signal after the analog-to-digital converter (ADC);
    \item Package boundary detection (PBD) error: due to correlator sensitivity of packet detector, the packet detection at the receiver introduces another time shift, resulting in a random phase offset to the raw CSI;
    \item Carrier frequency offset (CFO): The carrier frequencies between the transmitter and receiver cannot be perfectly synchronized. The CFO is initially compensated by a CFO corrector at the receiver. However, the compensation is incomplete due to the hardware imperfection, leaving residual CFO and leading to a phase offset over time;
    \item Phase offsets between RF chains: The CSI phases at different RF chains are not synchronized perfectly and it results in semi-deterministic phase offsets between RF chains;
\end{enumerate}
result in both amplitude and phase distortions to raw CSI measurements. 

\begin{figure}[t]
 \centering
 \subfloat[][raw CSI amp.]{
 \includegraphics[width=0.455\linewidth]{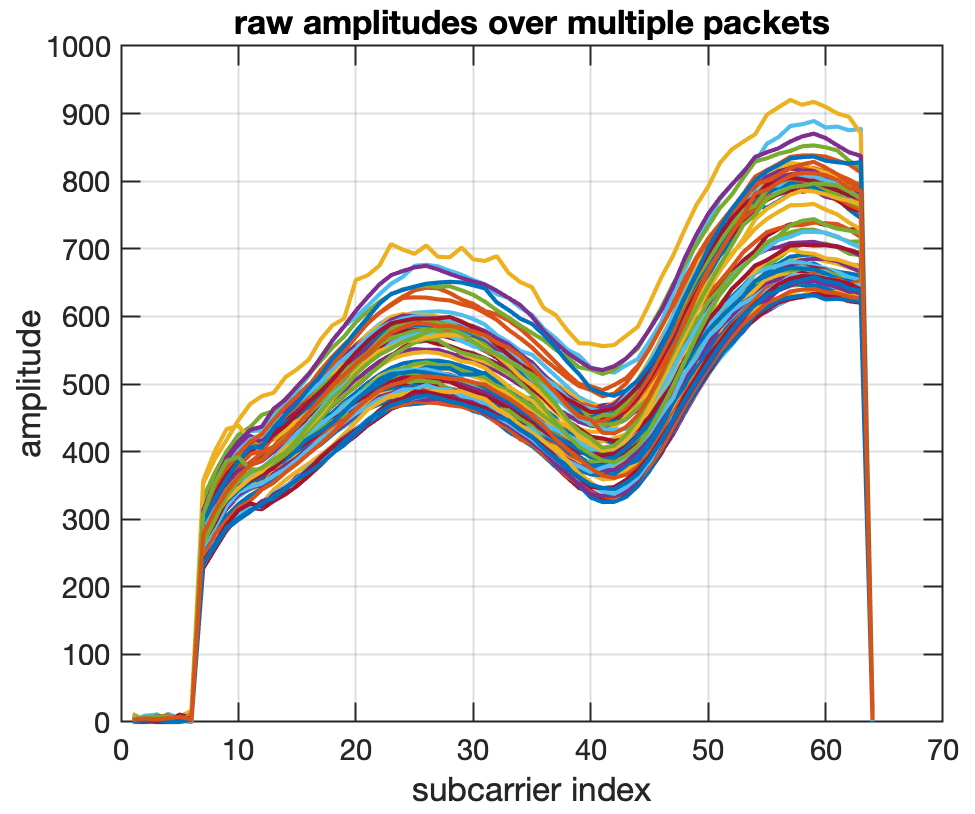}
 } \ \ 
 \subfloat[][calibrated CSI amp.]{
 \includegraphics[width=0.455\linewidth]{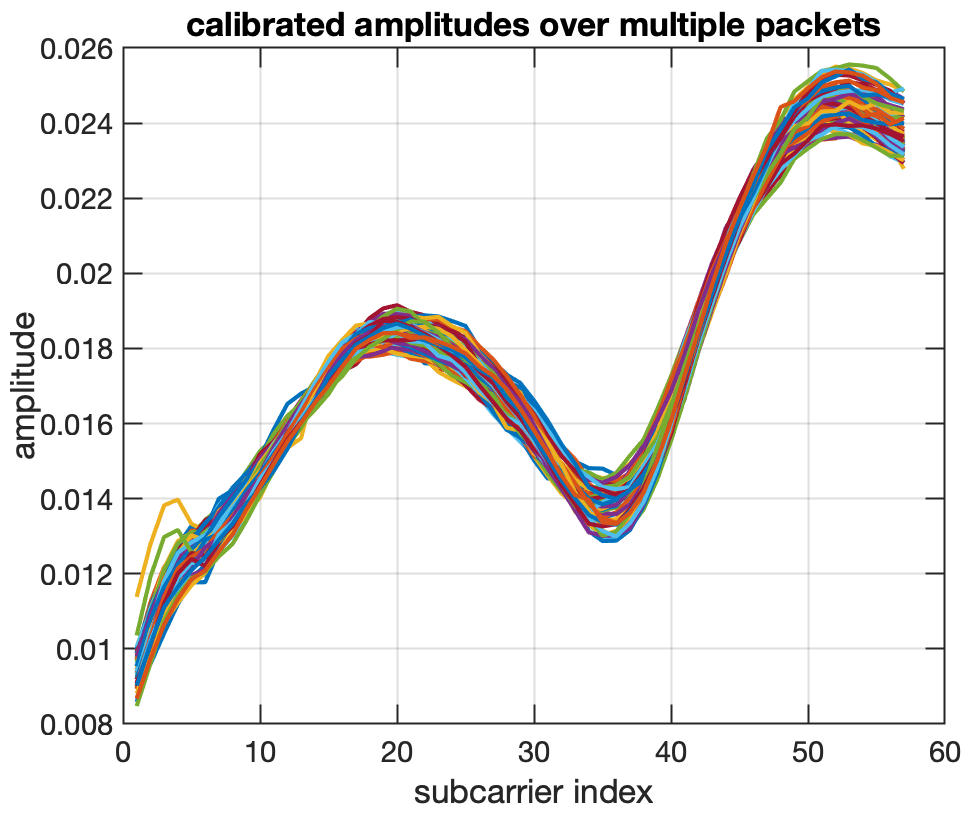}
 } \\ 
 \subfloat[][raw CSI phase]{
 \includegraphics[width=0.425\linewidth]{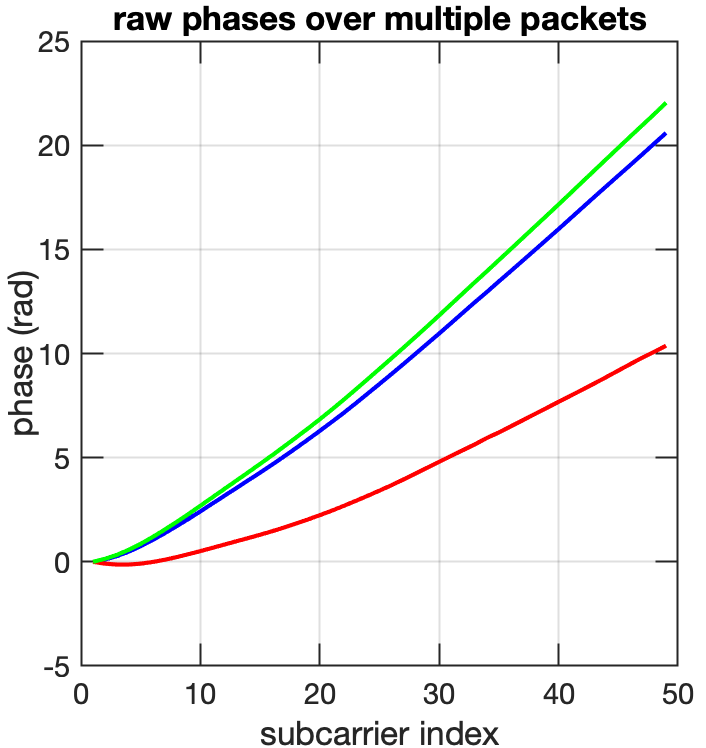}
 } \ \
 \subfloat[][calibrated CSI phase]{
 \includegraphics[width=0.435\linewidth]{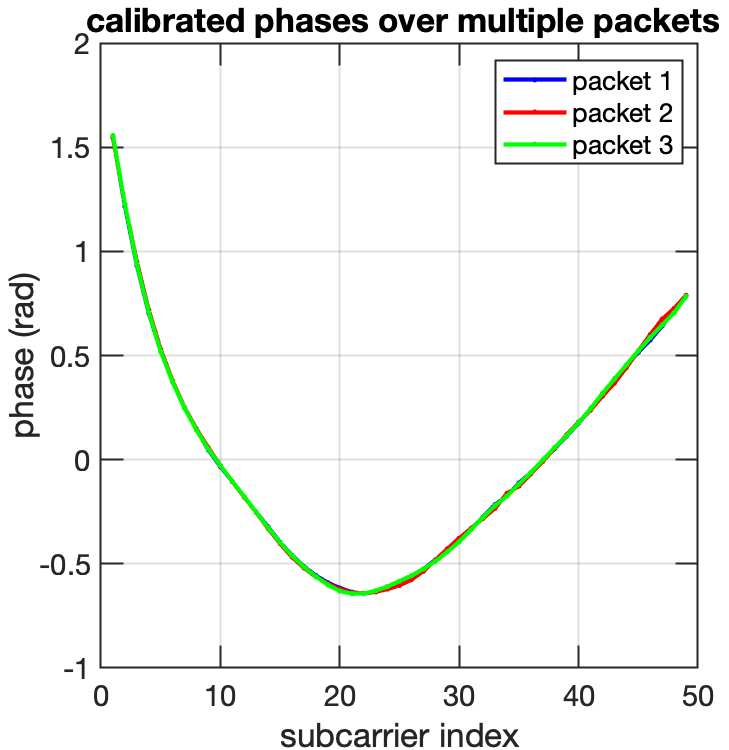}
 } \\
  \subfloat[][raw beam SNRs]{
 \includegraphics[width=0.55\linewidth]{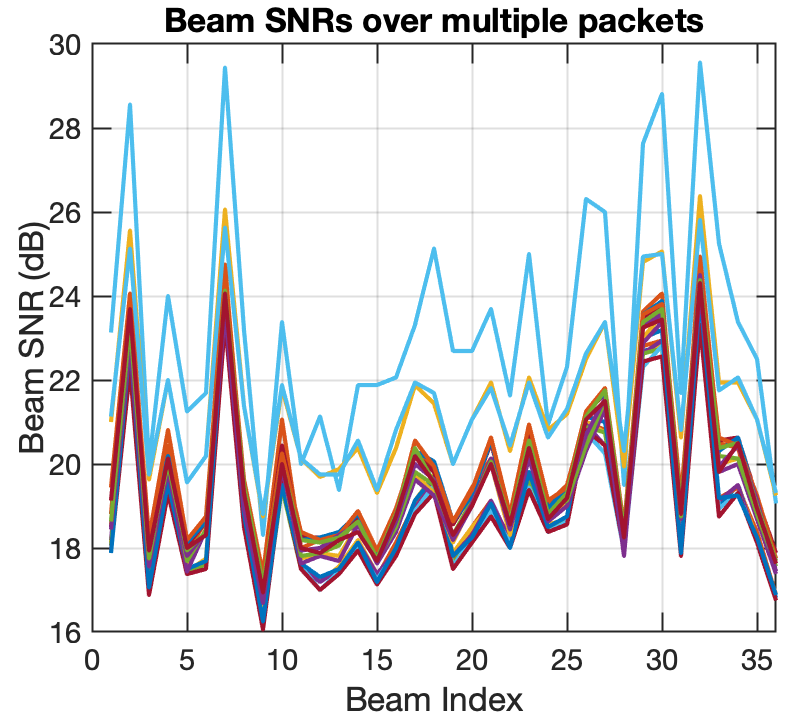}
 }
 \caption{Multi-band Wi-Fi measurement calibration and preprocessing. }
\label{fig_calbiration}	
\end{figure}

For Item 1), the CSI amplitude offsets in individual bands can be removed by an averaging or a standardization operation \cite{XieLi15}; see Fig.\ref{fig_calbiration} (a) and (b). For Items 2)-4), the phase offset over subcarriers and/or packets can be calibrated by either a linear phase fitting approach used in the SpotFi \cite[Algorithm 1]{KotaruJoshi15} or a multi-subband phase correction approach in \cite{XieLi15, ZhuoZhu17}. The compensated phases (with respect to the first subcarrier) of multiple packets are stable over time, as shown in Fig.\ref{fig_calbiration} (c) and (d). The combined amplitude and phase compensated CSI can serve as an indirect measurement of the time-of-flight (ToF) or delay power profile of the multi-path propagation. 

To further extract the angle (e.g., AoA) information across multiple antennas, one may need to compensate the initial phase offsets (of the phased-locked loops (PLLs)) between RF chains of Item 5) by connecting a coaxial cable between the transmitter and receiver \cite{ZhangHu20}. However, due to the semi-deterministic nature of this phase offset and the fixed amount of phase permutations ($4$ phase permutations for $3$ antennas), we skip this antenna calibration and treat the raw CSI across antennas as augmented CSI data. Moreover, we remove guard subcarriers as corresponding amplitudes are always zeros. 

On the other hand, we found that the beam SNRs as positive scalar values (so no need for the phase calibration) are much more consistent than the CSI measurement, except for a scaling effect over multiple packets; see Fig.~\ref{fig_calbiration} (e). As a result, the raw beam SNRs in the training data are augmented by multiplying them with a random scaling factor in $[0.9, 1.2]$ and then standardized over each beam index.

\begin{table}[t]
\caption{Number of beam SNR and CSI (in the bracket) samples for each class for pose recognition and occupancy sensing. } 
\centering 
\begin{tabular}{c | c | c | c | c} 
\hline\hline 
Class ID & Training-Pose & Test-Pose & Training-Occ. & Test-Occ.)  \\ 
\hline 
\hline
0 &  434 (6510) & 151  & 508 (7620) & 230  \\
1 & 499 (7485) & 149 & 717 (10755) & 290\\
2 & 325 (4875)  & 173 & 438 (6570)  & 189 \\   
3 & 347 (5205)  & 129  & 483 (7245)   & 184 \\
4 & 238 (3570) & 88 & 602 (9030) & 235  \\
5 & 314 (4710)  & 96 & 815 (12225) & 217   \\
6 &   272 (4080) & 119 & 535 (8025) & 200 \\
7 &   432 (6480)  & 135 & 608 (9120)  & 271 \\
\hline 
\end{tabular}
\label{table_sampleNumber} 
\end{table}

\subsection{Baseline Methods}
For a fair comparison, we consider a list of $4$ baseline methods with the proposed fusion network with granularity matching (GM) for both supervised learning and transfer learning: 
\begin{itemize}
 \item{\textbf{CSI-Only}} uses only the fine-grained CSI measurements as the input. The neural network follows the same architecture of the upper CSI encoder of Fig.~\ref{fig-featurePermutation} for supervised learning and the upper CSI branch (both encoder and decoder) of  Fig.~\ref{fig-fusion} for unsupervised fusion. The only difference is that the input is only the CSI and we only change the first convolutional layer configuration to reflect the change in the input dimension while keeping all other layers the same. 
  \item{\textbf{bSNR-Only}} uses only the mid-grained beam SNR (bSNR) measurements as the input. Similarly, we slightly change the first convolution layer of Fig.\ref{fig-featurePermutation} for supervised learning and the lower beam SNR branch (encoder and decoder) of Fig.~\ref{fig-fusion} for unsupervised fusion while keeping other layers the same. 
    \item{\textbf{Input-Fusion} (IF)} concatenates the CSI and beam SNR at the input. The concatenated input is then fed to the CSI encoder of Fig.~\ref{fig-featurePermutation} with the first convolution layer slightly changed and the CSI branch of Fig.~\ref{fig-fusion} for unsupervised fusion. 
    \item{\textbf{Feature-Fusion} (FF)} follows the architectures of Fig.\ref{fig-featurePermutation} and Fig.\ref{fig-fusion} by only fusing the outputs of the two encoders without the granularity matching.   
\end{itemize}


For performance evaluation, we use the confusion matrix $\Cbf$ as the criterion:
\begin{align}
\Cbf(i, j) = \frac{1}{T_j} \sum\limits_{t=1}^{T_j} \mathbbm{1}[\hat{l} = i],
\end{align}
where $i$ and $j$ are indices, respectively, for the estimated and true labels, and $T_j$ is the number of samples in the test dataset for the index $j$. In addition, $\hat{l}$ is the estimate by using the $t$th sample batch from the test data collected at the $j$th label. 

\begin{table}[t]
\caption{Task-specific supervised training: Average probability of successful classification as a function of the percentage of labeled  data $R$ to all training data $r+R$ (i.e., $R/(r+R)$)} 
\centering 
\begin{tabular}{c | c | c | c| c |c} 
\hline\hline 
$R/(r+R)$ & CSI &  bSNR & IF & FF & GM\\ 
\hline 
\text{All-Pose} & 74.8\% &  92.3\% & 89.5\% &  90.0\%  &  \textbf{94.4\%} \\
40\%                     & 64.0\% &  86.8\% & 84.2\% &  83.0\% &  \textbf{90.2\%}  \\
20\%                     & 57.1\% &  81.8\% & 83.2\% &  79.9\% &  \textbf{89.9\%} \\
10\%                     & 48.7\% &  77.9\% & 73.0\% &  77.8\% &  \textbf{80.0\%}  \\ 
\hline \hline 
\text{All-Occ.}        & 89.6\% &  88.5\% & \textbf{96.0\%} &  93.4\%  &  95.5\% \\
40\%                     & 80.2\% &  86.7\% & 92.7\% &  92.1\% &  \textbf{94.4\%}  \\
20\%                     & 76.4\% &  86.0\% & 91.2\% &  91.2\% &  \textbf{93.2\%} \\
10\%                     & 73.1\% &  76.0\% & 83.5\% &  82.6\% &  \textbf{86.8\%}  \\ 
\hline \hline 
\text{All-Loc.}        & 85.2\% &  92.5\% & \textbf{96.4\%} &  95.2\%  &  95.8\% \\
\hline \hline 
\end{tabular}
\label{table_SL}
\end{table}

\begin{figure}[t]
 \centering
 \subfloat[][CSI]{
 \includegraphics[width=0.23\linewidth]{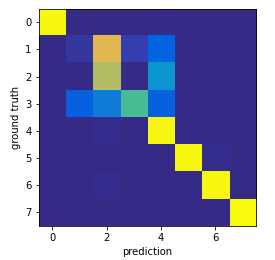}
 }
 \subfloat[][bSNR]{
 \includegraphics[width=0.23\linewidth]{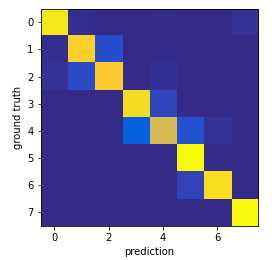}
 } 
 \subfloat[][IF]{
 \includegraphics[width=0.225\linewidth]{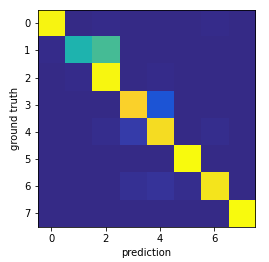}
 }
 \subfloat[][GM]{
 \includegraphics[width=0.22\linewidth]{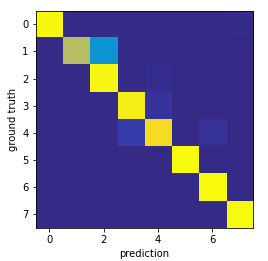}
 } \\
  \subfloat[][CSI]{
 \includegraphics[width=0.23\linewidth]{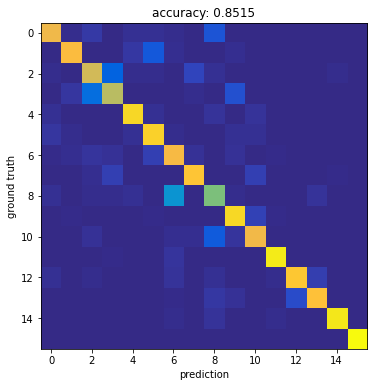}
 }
 \subfloat[][bSNR]{
 \includegraphics[width=0.23\linewidth]{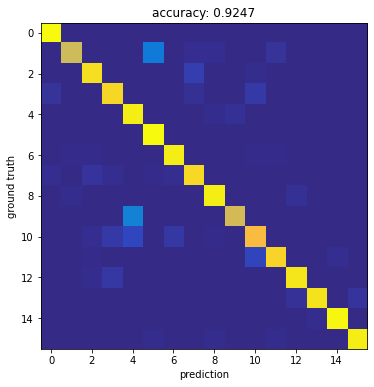}
 } 
 \subfloat[][IF]{
 \includegraphics[width=0.225\linewidth]{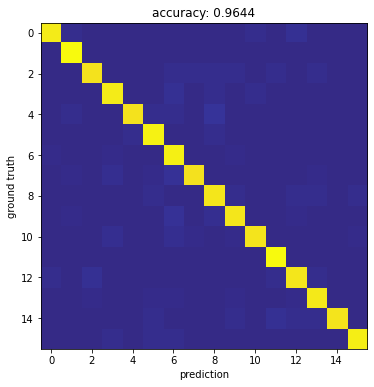}
 }
 \subfloat[][GM]{
 \includegraphics[width=0.22\linewidth]{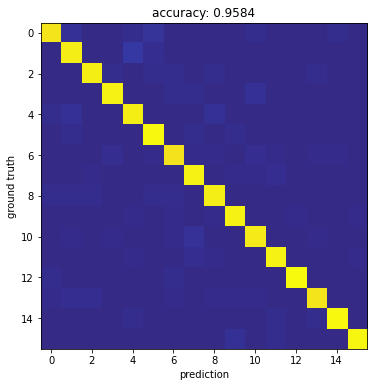}
 }
 \caption{Confusion matrices for pose recognition (top row) and indoor localization (bottom row) when all training data are labeled. }
\label{fig_SL_CM}	
\end{figure}

\subsection{Supervised Classification}
\label{perfClassification}
In Table~\ref{table_SL}, we first present the results for the task-specific supervised training for the $5$ considered methods for  pose recognition, occupancy sensing, and indoor localization. When all training data in Table~\ref{table_sampleNumber} are labeled (e.g., the two rows of All-Pose and All-Occ.), all methods give reasonably good performance except the CSI-only method for the pose recognition. For the pose recognition, it seems that the CSI-only measurements provide less distinct features compared with the beam SNRs as one can see the classification accuracy drops from $92.3\%$ to $74.8\%$. This makes the fusion between a ``good" measurement and a ``bad" measurement much more challenging. Indeed, it is noted that, for the pose recognition, a simple input fusion between the CSI and beam SNR reduces the classification accuracy from $92.3\%$ for the beam SNR-only to $89.5\%$. With the FF, the accuracy is slightly better at $90.0\%$ but still less than that of the beam SNR-only method. Finally, with the GM that takes into account the feature granularity correspondence, it can achieve better accuracy compared with the beam SNR-only method. Corresponding confusion matrices are shown in the top row of Fig.~\ref{fig_SL_CM}. 

For the occupancy sensing application, the CSI and beam SNR measurements give similar accuracy levels around $89\%$ which indicates the similar quality of both measurements. In this case, the simple IF gives the highest performance at $96.0\%$ while the FF and GM provide an accuracy of $93.4\%$ and $95.5\%$, respectively. Overall, it is seen that, when both measurements are of good quality, all three fusion methods can boost the performance of individual measurements by a margin of $5\% - 8\%$. 

For indoor localization, all methods give relatively good results. All three fusion methods can further improve the accuracy to more than $95\%$ compared with $92.5\%$ for the beam SNR-only measurement and $82.5\%$ for the CSI-only measurements. Corresponding confusion matrices are shown in the bottom row of Fig.~\ref{fig_SL_CM}.

\begin{table}[t]
\caption{Unsupervised fusion+transfer learning: Average probability of successful classification as a function of the percentage of labeled  data $R$ to all training data $r+R$ (i.e., $R/(r+R)$)} 
\centering 
\begin{tabular}{c | c | c | c| c |c} 
\hline\hline 
$R/(r+R)$ & CSI &  bSNR & IF & FF & GM\\ 
\hline 
\text{All-Pose}      & 75.4\% &  87.5\% & 86.0\% &  89.3\%  &  \textbf{91.2\%} \\
40\%                     & 72.3\% &  84.6\% & 82.6\% &  86.8\% &  \textbf{87.5\%}  \\
20\%                     & 74.0\% &  85.9\% & 82.1\% &  85.0\% &  \textbf{86.5\%} \\
10\%                     & 75.9\% &  85.3\% & 82.0\% &  86.5\% &  \textbf{88.0\%}  \\ 
\hline \hline 
\text{All-Occ.}        & 90.1\% &  75.2\% & 84.0\% &  88.5\%  &  \textbf{92.5\%} \\
40\%                     & 89.6\% &  71.8\% & 82.2\% &  88.0\% &  \textbf{90.3\%}  \\
20\%                     & \textbf{88.4\%} &  71.9\% & 82.9\% &  87.4\% &  {86.6\%} \\
10\%                     & 78.2\% &  76.7\% & 83.8\% &  88.0\% &  \textbf{89.2\%}  \\ 
\hline 
\end{tabular}
\label{table_UL}
\end{table}

\begin{figure}[t]
     \begin{center}
	 \includegraphics[width=0.45\textwidth]{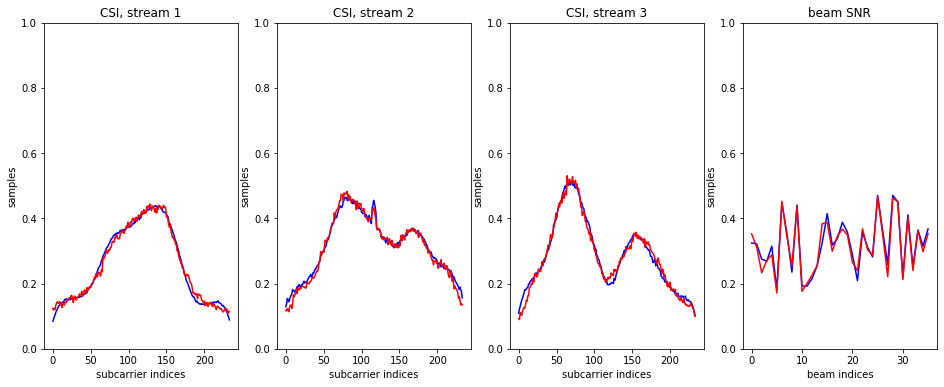} 
	    \end{center}
    \caption{Waveform comparison between the input to the autoencoder-based unsupervised fusion and its output in  Fig.~\ref{fig-fusion}. Blue: original input waveforms; Red: reconstructed output waveforms from the same fused latent space. }
   \label{fig-waveform}
\end{figure}

\subsection{Unsupervised Fusion and Transfer Learning}
\label{perfClassification}
Meanwhile, we provide the results for unsupervised fusion with transfer learning for the $5$ considered methods for both pose and occupancy sensing. The results are summarized in Table~\ref{table_UL}. Fig.~\ref{fig-waveform} illustrates the comparison between the original input waveforms, i.e., $3$ spatial streams of CSI and $1$ set of beam SNRs,  and the reconstructed output waveforms from the same fused latent space using the autoencoder-based unsupervised fusion network of Fig.~\ref{fig-fusion}. 

When all training data are labeled (e.g., the two rows of All-Pose and All-Occ.), the GM can provide the best classification results compared with individual measurement-based methods and the two other fusion methods. It is also interesting to compare the results with their counterparts in Table~\ref{table_SL}. It is expected that the classification accuracy of the unsupervised fusion can be relatively worse than the supervised training when all training data are labeled. The results show that our model can learn well with unlabelled data, and predict better than its corresponding supervised model with limited labeled data.

\begin{figure}[t]
 \centering
   \subfloat[][]{
 \includegraphics[width=0.23\linewidth]{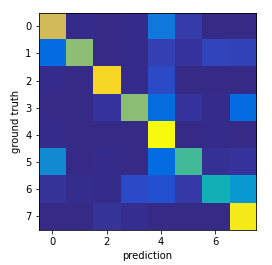}
 }
 \subfloat[][]{
 \includegraphics[width=0.23\linewidth]{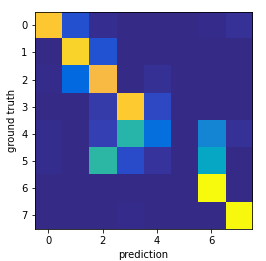}
 } 
 \subfloat[][]{
 \includegraphics[width=0.225\linewidth]{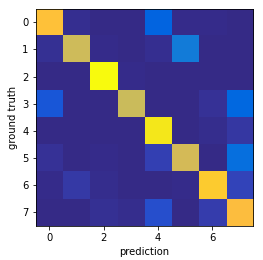}
 }
 \subfloat[][]{
 \includegraphics[width=0.23\linewidth]{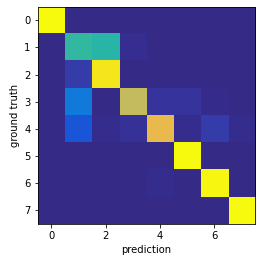}
 } \\
 \subfloat[][CSI]{
 \includegraphics[width=0.23\linewidth]{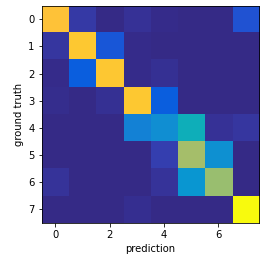}
 }
 \subfloat[][bSNR]{
 \includegraphics[width=0.23\linewidth]{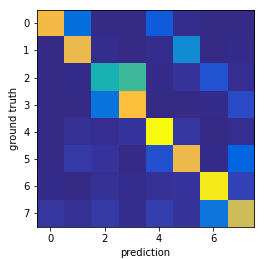}
 } 
 \subfloat[][IF]{
 \includegraphics[width=0.225\linewidth]{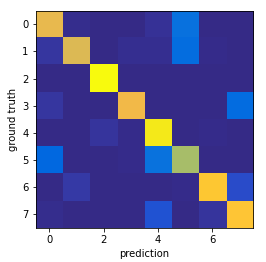}
 }
 \subfloat[][GM]{
 \includegraphics[width=0.23\linewidth]{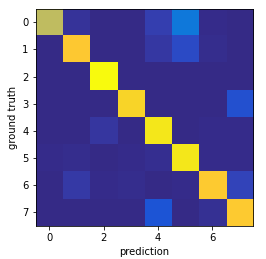}
 }
 \caption{Confusion matrices for occupancy sensing with supervised training (top row)  and unsupervised fusion and transfer learning (bottom row) when only $10\%$ training data are labeled.}
\label{fig_UL_CM}	
\end{figure}

\begin{figure}[t]
     \begin{center}
	 \includegraphics[width=0.4\textwidth]{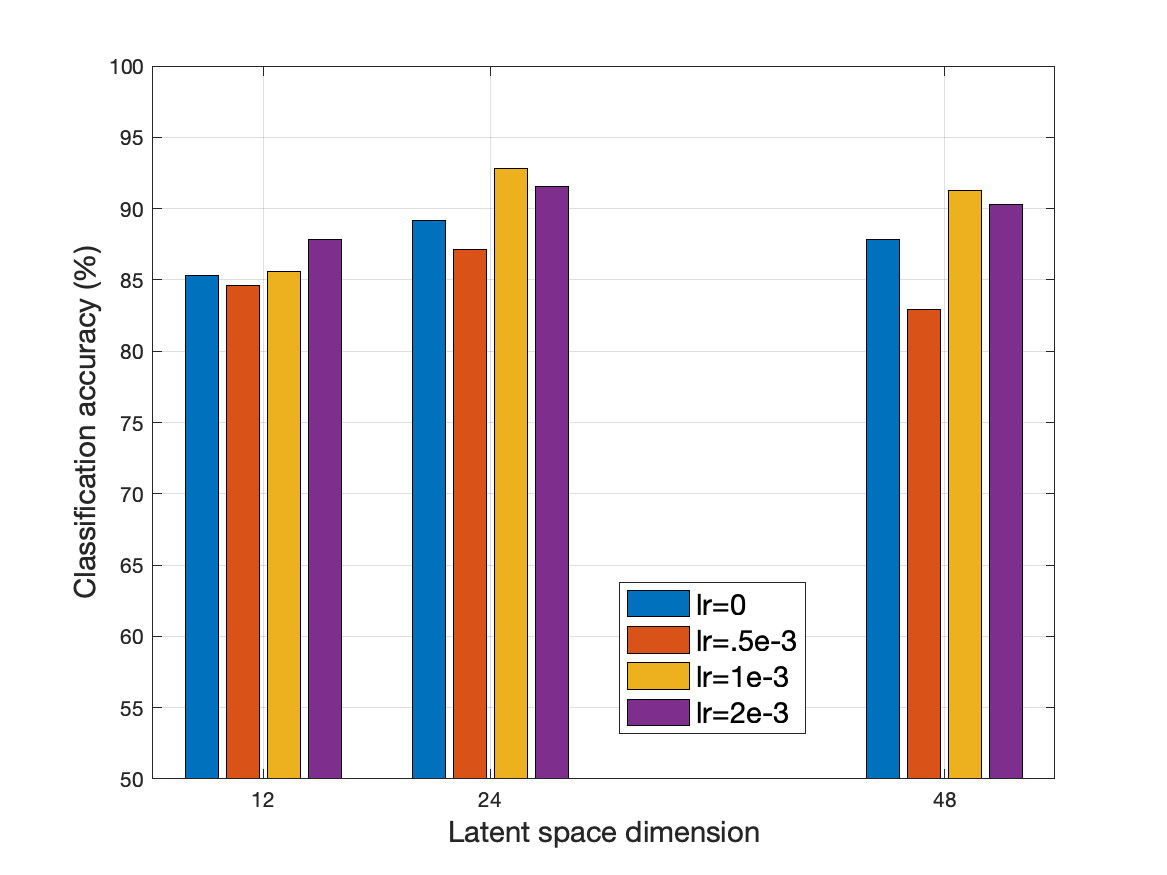} 
	    \end{center}
    \caption{Impact of latent space dimension $d$ and the learning rate ($lr$) used for finetuning of the fusion block.}
   \label{fig-latent}
\end{figure}

\subsection{Impact of Labeled Sample Size}
We also performed a comprehensive evaluation on the impact of the percentage of the labeled data to the whole training data, namely $R/(r+R)$. In Table~\ref{table_SL} and Table~\ref{table_UL}, we list the classification accuracy for all considered methods when the percentage of labeled data reduces to $40\%$, $20\%$, and $10\%$. 

For the supervised training, one can notice that the performance gradually decreases as the number of labeled data is less. With only $10\%$ labeled training data, the fusion-based methods give the classification accuracy in between $73\%$ and $86.8\%$ for both pose recognition and occupancy sensing. Corresponding confusion matrices for the occupancy sensing are shown in the top row of Fig.~\ref{fig_UL_CM}. 

For the unsupervised fusion and transfer learning in Table~\ref{table_UL}, one can also conclude that the performance degrades as the number of labeled data reduces. However, compared with the results in Table~\ref{table_SL}, the amount of performance degradation appears to be smaller in Table~\ref{table_UL}, potentially due to the use of the autoencoder-based unsupervised fusion. More notably, when the percentage of labeled training data reduces to $10\%$, the unsupervised fusion and transfer learning can achieve better results than their supervised learning counterparts. For a better visual comparison, we also show the confusion matrices in the bottom row of Fig.~\ref{fig_UL_CM} by using the unsupervised fusion and transfer learning. 
\begin{figure}[t]
 \centering
 \includegraphics[width=.55\linewidth]{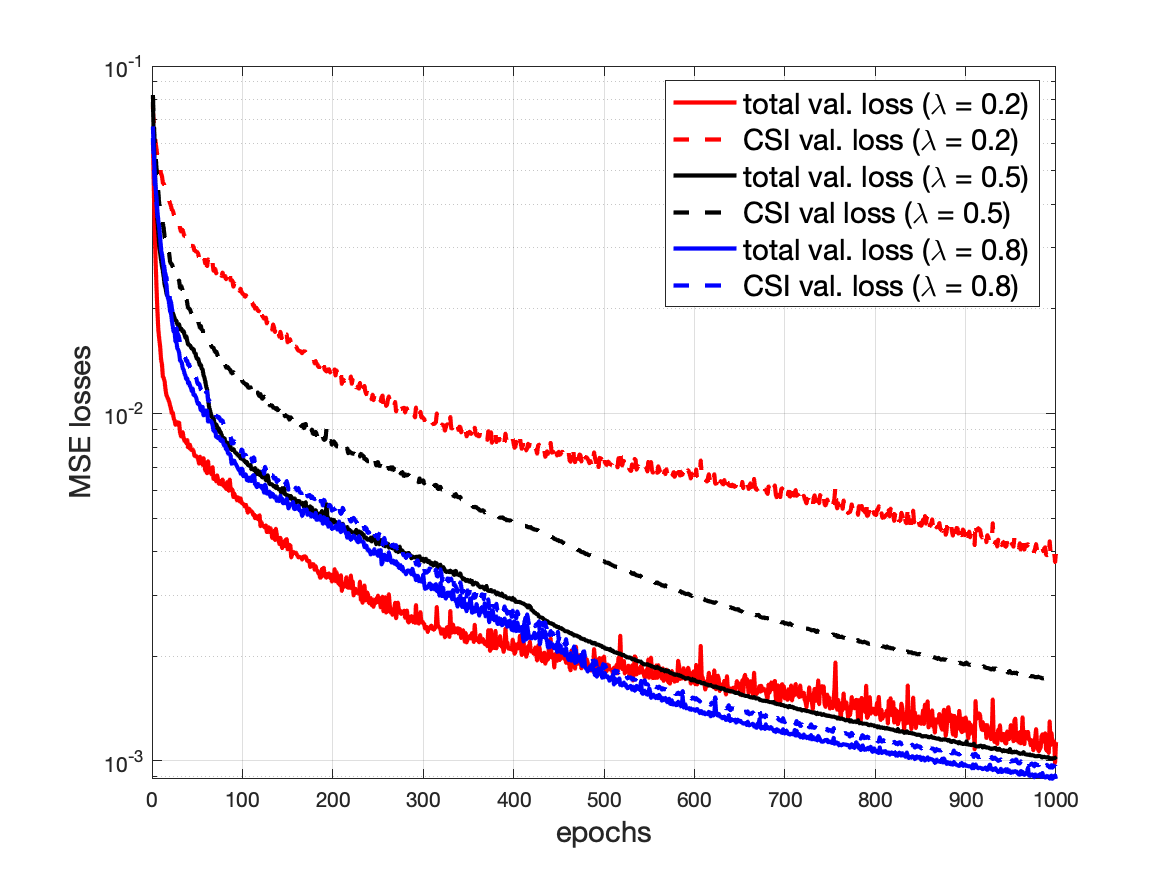}
 \caption{\textcolor{black}{Learning error trajectories in terms of MSEs over epochs.}}
 \label{fig-epoch}
 \vspace{-0.2in}
\end{figure}


\begin{figure*}[t]
 \centering
   \subfloat[][]{
 \includegraphics[width=0.24\linewidth]{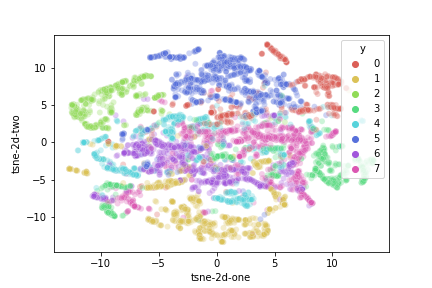}
 }
 \subfloat[][]{
 \includegraphics[width=0.24\linewidth]{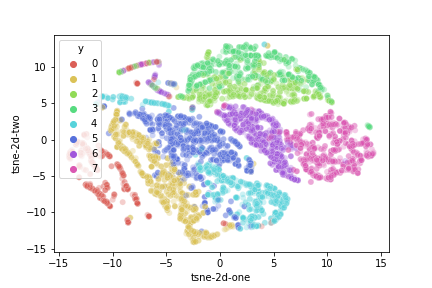}
 } 
 \subfloat[][]{
 \includegraphics[width=0.24\linewidth]{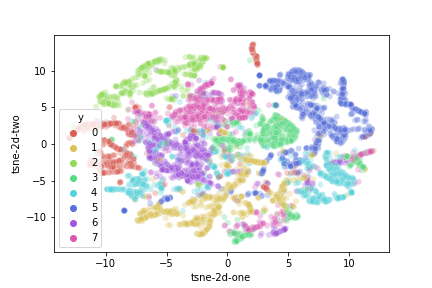}
 }
 \subfloat[][]{
 \includegraphics[width=0.24\linewidth]{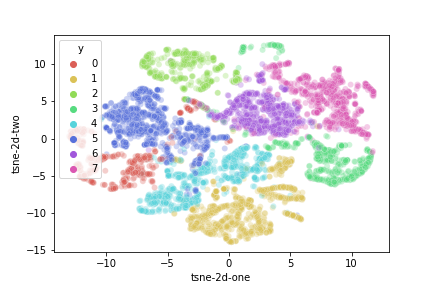}
 } 
 \caption{Visualization of latent space clusters using t-SNE for a) CSI; b) beam SNR; c) input fusion; and d) granularity matching. }
\label{fig_tsne}	
\vspace{-0.2in}
\end{figure*}

\subsection{Impact of Latent Space Dimension and Finetuning Rate}
Following the above discussion on the transfer learning when only $10\%$ labeled training data are available, we further consider the impact of the latent space dimension $d$ and the learning rate ($lr$) used for fine-tuning of the fusion block. Particularly, we vary the dimension of the fused feature from $d=12$, $d=24$ to $d=48$ and use $4$ learning rates, i.e., $lr=\{0, .5e-3, 1e-3, 2e-3\}$, for fine-tuning the fusion weight in \eqref{lw}. As shown in Fig.~\ref{fig-latent}, the fine-tuning (with $lr>0$) appears to always improve the classification accuracy for all three cases of latent space dimensions, although the best learning rate varies from one case to another. For instance, $lr=2e-3$ gives the best performance when $d=12$, while $lr=1e-3$ is the best for $d=24$ and $d=48$. 
 
Meanwhile, if we compare the results over $d$, the best average performance is given by the latent space dimension $d=24$ with $d=48$ comes the second. $d=12$ gives the worse performance among the three considered latent space dimensions. One plausible explanation for $d=24$ being the best is that the latent space dimension of $d=24$ is in between the input dimension of the beam SNR $M=36$ and the output dimension of $N=8$, while $d=12$ is limited in terms of capacity and $d=48$ is too large with respect to the input dimension. 

\subsection{Learning Trajectories}
Fig.~\ref{fig-epoch} shows the learning error trajectories in terms of the validation loss as a function of epochs for the autoencoder-based unsupervised fusion. More specifically, the validation loss is computed using the validation dataset according to the weighted MSE of \eqref{wmse} with three choices of the weight $\lambda$. We also include the corresponding CSI-based MSE component in the figure. It is seen that, by placing larger weights on the CSI terms, the total validation loss converges to smaller values (blue curves versus red curves), while smaller weights (particularly $\lambda=0.2$) leads to quicker MSE reduction at the beginning, e.g., for the first $400$ epochs. 

We then used the three pretrained encoder and fusion networks and applied the transfer learning with $10\%$ labeled data. The classification accuracy improves from $88.4\%$ when $\lambda=0.2$, $89.2\%$ when $\lambda=0.5$ (also reported in Table~\ref{table_UL}), to $90.3\%$ when $\lambda=0.8$. 


\subsection{Feature Visualization}
Finally, we use the tool of t-distributed Stochastic Neighbor Embedding (t-SNE) \cite{vanDerMaaten08} to visualize high-dimensional latent space in the autoencoder of  Fig.~\ref{fig-fusion} (a). t-SNE converts similarities among data points to joint probabilities and tries to minimize the Kullback-Leibler divergence between the joint probabilities of the low-dimensional embedding and the high-dimensional data. 

Fig.~\ref{fig_tsne} shows the t-SNE results when the latent space dimension is $d=24$. We include a) CSI-only autoencoder, b) beam SNR-only autoencoder, c) input fusion, and d) granularity matching to illustrate the cluster separation in the corresponding latent space. It is clear to see that the CSI measurements are less separated in the latent space compared with the beam SNR. The best clustering separation appears to be achieved by the GM in Fig.~\ref{fig_tsne} (d). 

\section{Conclusion}
\label{sec:conclusion}
This paper has demonstrated that, by fusing the sub-6 GHz CSI measurements and mmWave 60-GHz beam SNR measurements, one can achieve around 5\% gain in accuracy more robust performance for various sensing tasks. Particularly, we considered both supervised classification with granularity matching and an unsupervised autoencoder-based fusion with transfer learning to make better use of unlabeled training data and mitigate the need for massive fingerprinting data, while it is able to maintain its performance requiring around 40\% data labeled. We also conducted comprehensive data collection and performance validation in terms of classification accuracy and various impact factors.

\bibliographystyle{IEEEtran}
\bibliography{bib_localization}  

\end{document}